%% file: main.tex
\documentclass[conference]{IEEEtran}
\IEEEoverridecommandlockouts
\pdfoutput=1
\usepackage{cite}
\usepackage{amsmath,amssymb,amsfonts}
\usepackage{algorithm}
\usepackage{algpseudocode}
\usepackage{graphicx}
\usepackage{textcomp}
\usepackage{xcolor}
\usepackage{url}
\usepackage{booktabs}
\usepackage{caption}
\usepackage{subcaption}
\usepackage{multirow}
\usepackage[frozencache=true, cachedir=.]{minted}
\input{default.pygstyle}

\usepackage[nogroupskip,nonumberlist,nopostdot]{glossaries}
\usepackage{glossary-mcols}
\input{src/acronyms}

\def\BibTeX{{\rm B\kern-.05em{\sc i\kern-.025em b}\kern-.08em
    T\kern-.1667em\lower.7ex\hbox{E}\kern-.125emX}}

\def\Cpp{C\texttt{++}}

\makeatletter
\newcommand{\linebreakand}{%
  \end{@IEEEauthorhalign}
  \hfill\mbox{}\par
  \mbox{}\hfill\begin{@IEEEauthorhalign}
}
\makeatother

\begin{document}

% \title{Morpheus: Fast cross-platform SpMV\\on emerging architectures}
% \title{Performance portable cross-platform SpMV\\on emerging architectures}
\title{Morpheus unleashed: Fast cross-platform\\SpMV on emerging architectures}

\author{%
\IEEEauthorblockN{Christodoulos Stylianou}
\IEEEauthorblockA{\textit{EPCC, The University of Edinburgh} \\
Edinburgh, United Kingdom \\
c.stylianou@ed.ac.uk}
\and
\IEEEauthorblockN{Mark Klaisoongnoen}
\IEEEauthorblockA{\textit{EPCC, The University of Edinburgh} \\
Edinburgh, United Kingdom \\
mark.klaisoongnoen@ed.ac.uk}
\and
\IEEEauthorblockN{Ricardo Jesus}
\IEEEauthorblockA{\textit{EPCC, The University of Edinburgh} \\
Edinburgh, United Kingdom \\
rjj@ed.ac.uk}
\linebreakand %
\IEEEauthorblockN{Nick Brown}
\IEEEauthorblockA{\textit{EPCC, The University of Edinburgh} \\
Edinburgh, United Kingdom \\
n.brown@epcc.ed.ac.uk}
\and
\IEEEauthorblockN{Mich\`{e}le Weiland}
\IEEEauthorblockA{\textit{EPCC, The University of Edinburgh}\\
Edinburgh, United Kingdom \\
m.weiland@epcc.ed.ac.uk}
}

\maketitle

\begin{abstract}
Sparse matrices and linear algebra are at the heart of scientific simulations. Over the years, more than 70 sparse matrix storage formats have been developed, targeting a wide range of hardware architectures and matrix types, each of which exploit the particular strengths of an architecture, or the specific sparsity patterns of the matrices.

In this work, we explore the suitability of storage formats such as COO, CSR and DIA for emerging architectures such as AArch64 CPUs and \glspl{fpga}. In addition, we detail hardware-specific optimisations to these targets and evaluate the potential of each contribution to be integrated into \emph{Morpheus}, a modern library that provides an abstraction of sparse matrices (currently) across x86 CPUs and NVIDIA/AMD GPUs. Finally, we validate our work by comparing the performance of the Morpheus-enabled HPCG benchmark against vendor-optimised implementations.
\end{abstract}

\begin{IEEEkeywords}
sparse matrix storage formats, AArch64, FPGA, performance portability, productivity
\end{IEEEkeywords}

\input{src/introduction}
\input{src/background}
\input{src/implementation}

\input{src/experiments}

\input{src/related_work}
\input{src/conclusions}
\input{src/acknowledgments}

\bibliographystyle{IEEEtran}
% \bibliography{bibliography/bibs}
\bibliography{bibliography/main}

\end{document}

%% file: src/acronyms.tex
\newacronym{hpc}{HPC}{High Performance Computing}
\newacronym{fea}{FEA}{Finite Element Analysis}
\newacronym{fdm}{FDM}{Finite Differences Method}
\newacronym{fem}{FEM}{Finite Element Method}
\newacronym{fvm}{FVM}{Finite Volume Method}
\newacronym{pde}{PDE}{Partial Differential Equation}
\newacronym{ai}{AI}{Artificial Intelligence}
\newacronym{ksp}{KSP}{Krylov Subspace Methods}
\newacronym{dof}{DOF}{Degrees of Freedom}
\newacronym{ebe}{EBE}{Element-by-element}
\newacronym{mp}{MP}{Mixed-precision}
\newacronym{gpu}{GPU}{Graphics Processing Unit}
\newacronym{mpi}{MPI}{Message-passing Interface}
\newacronym{cpu}{CPU}{Central Processing Unit}
\newacronym{simd}{SIMD}{Single Instruction Multiple Data}
\newacronym{cg}{CG}{Conjugate Gradient}
\newacronym{pcg}{PCG}{Preconditioned Conjugate Gradient}
\newacronym{minres}{MINRES}{Minimal Residual Method}
\newacronym{gmres}{GMRES}{Generalised Minimal Residual Method}
\newacronym{bicg}{BiCG}{Biconjugate Gradient Method}
\newacronym{acg}{ACG}{Adaptive Conjugate Gradient}
\newacronym{spmv}{SpMV}{Sparse Matrix-Vector Multiplication}
\newacronym{spgemm}{SpGEMM}{Sparse Matrix-Matrix Multiplication}
\newacronym{amg}{AMG}{Algebraic Multi-grid}
\newacronym{gmg}{GMG}{Geometric Multi-grid}
\newacronym{dt}{DT}{Decision Tree}
\newacronym{cnn}{CNN}{Convolutional Neural Network}
\newacronym{ml}{ML}{Machine Learning}
\newacronym{dl}{DL}{Deep Learning}
\newacronym{sor}{SOR}{Successive Over-relaxation}
\newacronym{arm}{ARM}{Acorn RISC Machine}

\newacronym{coo}{COO}{Coordinate}
\newacronym{csr}{CSR}{Compressed Sparse Row}
\newacronym{dia}{DIA}{Diagonal}
\newacronym{ell}{ELL}{ELLPACK}
\newacronym{hyb}{HYB}{Hybrid}
\newacronym{hdc}{HDC}{Hybrid DIA/CSR}
\newacronym{hec}{HEC}{Hybrid ELL/CSR}

\newacronym{vtable}{vtable}{virtual function table}
\newacronym{hpcg-bench}{HPCG}{High Performance Conjugate Gradients}
\newacronym{hpl-bench}{HPL}{High Performance LINPACK}
\newacronym{hbm}{HBM2}{High Bandwidth Memory}

\newacronym{sve}{SVE}{Scalable Vector
Extension}
\newacronym{armpl-acr}{ArmPL}{Arm Performance Libraries}
\newacronym{fpga}{FPGA}{Field Programmable Gate Array}
\newacronym{xrt}{XRT}{Xilinx Runtime Library}

%% file: src/introduction.tex
\section{Introduction}

Since their inception, sparse matrices have become the centrepiece of many applications in science and engineering. Their ability to efficiently store the non-zero values of a matrix reduces the memory footprint of the matrix and eliminates redundant computations, allowing for larger problems to be tackled. More than 70 sparse matrix storage formats (i.e data structures) have been introduced over the years~\cite{spmv_gpu_review}, each leveraging different properties of the matrix or target hardware architecture to achieve better performance. Many iterative methods for solving large-scale linear systems and eigenvalue problems, which often arise in a variety of scientific and engineering applications, consist of many \glspl{spmv} operations that often dominate the applications' runtime. Literature shows that no single format can perform optimally across all kinds of matrices or hardware~\cite{spmv_gpu_review,cnn_spmv,csr5,axt,sell_c_sigma}, with interest on optimising \gls{spmv} being renewed every time new platforms emerge.

\emph{Morpheus}~\cite{morpheus-p3hpc} is a \Cpp{} library that provides an abstraction of sparse matrices, allowing for efficient and transparent switching of sparse matrix storage formats at runtime  across traditional backends such as x86 CPUs and NVIDIA/AMD GPUs. By dynamically adapting the underlying sparse matrix data-structure to optimally suit an operation, target architecture, or sparsity pattern of the matrix, \emph{Morpheus} enables new optimisation opportunities and thus increased performance~\cite{morpheus-p3hpc}. At the time of writing, \emph{Morpheus} supports three core formats: \gls{coo}, \gls{csr} and \gls{dia}.

In this work, we explore the performance of \gls{spmv} on different formats across two non-traditional \gls{hpc} architectures, AArch64 CPUs and \glspl{fpga}, both currently gaining traction in the area of \gls{hpc}. In addition, we exploit specific hardware features of each target, such as the \gls{sve} on AArch64, to optimise the kernels, and investigate the challenges in integrating the optimisations in \emph{Morpheus}. 
In summary, our contributions are:
\begin{itemize}
    \item We discuss how we incorporated the \gls{armpl-acr} into Morpheus for AArch64 targets, thus making the performance of \gls{armpl-acr} available to Morpheus users ``out-of-the-box'' for \gls{coo} and \gls{csr}. 
    \item To work around \gls{armpl-acr}'s lack of DIA support and to enable a better assessment of ArmPL's performance, we develop \gls{sve}-optimised \gls{spmv} routines for COO, CSR and DIA matrices.
    \item For over 2100 matrices available in the SuiteSparse\cite{suitesparse} collection, we demonstrate the performance of the \gls{spmv} kernels on an A64FX-based (AArch64) HPE Apollo 80 cluster.
    \item We evaluate our efforts (i.e.\ incorporating \gls{armpl-acr} and our \gls{sve}-optimised routines into Morpheus) on HPCG\cite{hpcg}, comparing our Morpheus-based HPCG implementation~\cite{morpheus-hpcg} against the Arm-optimised version~\cite{arm-hpcg} on the HPE Apollo 80 system.%Note that the techniques applied in optimising the \gls{dia} format are transferable to other formats; therefore, our approach acts as a proof-of-concept.
    \item We provide implementations of the \gls{spmv} kernel on \glspl{fpga} for each of the three formats. Performance is evaluated on the SuiteSparse set and the challenges in porting the implementations in \emph{Morpheus} are described. 
\end{itemize}
% Furthermore, we investigate the integration of the optimised routines in \emph{Morpheus} and benchmark their performance against vendor provided versions of the \gls{hpcg-bench}\cite{hpcg} benchmark on systems such as the Edinburgh-based HPE Cray EX ARCHER2 and the Bristol-based HPE Apollo 80 partition of Isambard. In the sections below we detail the specific work undertaken for each of the aforementioned targets.

%% file: src/background.tex
\section{Motivation}
New storage formats are proposed every time new architectures emerge, aiming to exploit the new characteristics and features of the new hardware. Even-though more than 70 formats are available, no single format performs best across different hardware, operations, and sparsity patterns. As a result, being able to switch formats dynamically offers new opportunities for optimisation and increased performance. The adoption of new formats can be a tedious process as this requires significant changes in the source code. Libraries such as \emph{PETSc}\cite{petsc-web-page}, \emph{GINKGO}\cite{ginkgo} and \emph{Morpheus}\cite{morpheus-p3hpc} offer multiple formats through various abstractions, enabling users can switch to different formats at runtime. In other words, users can take advantage of new formats added in the library with minimal source code changes, thereby easing their development effort.

Nevertheless, when it comes to the adoption of new hardware, libraries might require major changes in their interface, especially when the hardware utilises a new programming model that the library does not yet integrate. For example, when GPUs emerged in \gls{hpc}, libraries had to evolve to manage the existence of two mostly independent memory spaces between the CPU and GPU and to support the accelerator model, where CPUs offload work to the GPUs.
For software to remain performant throughout the life-cycle of an hardware architecture (and beyond), we need to ensure that it can adapt to the requirements that are imposed by the current and future iterations of hardware.

In this work, we consider AArch64 CPUs and \glspl{fpga}, two very different architectures that are currently gaining traction in \gls{hpc}. We investigate their performance in sparse linear operations, such as \gls{spmv}, as well as the challenges involved in integrating them in \emph{Morpheus}.
The integration of code in \emph{Morpheus} for the aforementioned targets presents different challenges and requires different integration approaches, making the candidates representative for the integration of other architectures.

\section{Background}
\subsection{Emerging Architectures}
%\textcolor{red}{TODO: General background of emerging HW leading the discussion to AArch64 and FPGAs }

The two emerging architectures that we propose to integrate with the \emph{Morpheus} libraries are AArch64 CPUs and FPGAs. Through the process of porting the SpMV routines for the three core \emph{Morpheus} storage formats (COO, CSR and DIA) to AArch64 CPUs and to FPGAs, we explore the challenges around performance optimisation and performance portability for integration into the Morpheus library.

\subsubsection{AArch64 CPUs}
Despite being somewhat newcomers on the HPC scene, Arm CPUs have
already proved to be extremely competitive against the more traditional x86
processors~\cite{jackson2020investigating, nakao2021performance,
jesus2021vectorising, masoudkoleini2022graviton3}. One of the key enablers for
the high-performance of modern-day AArch64 CPUs is \gls{sve}~\cite{sve, stephens2017arm}, an advanced architecture
extension for \gls{simd} processing that features
long (scalable) vectors, gather-load and scatter-store instructions, and
per-lane predication. These features make \gls{sve} an excellent architecture to
implement fast \gls{spmv} computations.

To accelerate the adoption of Arm-based hardware in HPC, Arm has developed the \gls{armpl-acr}~\cite{armpl}, a set of core routines for high-performance computing applications
optimised for AArch64 processors. \gls{armpl-acr} consists of BLAS, LAPACK, FFT, Sparse,
libamath (subset of libm) and libastring (subset of libc for strings) routines
for both single-threaded and OpenMP multi-threaded processing. The sparse
linear algebra routines provided by \gls{armpl-acr} support high-performance \gls{spmv} on dense,
CSR, CSC, COO and BSR matrices. However, as of version 23.04 (the most recent
at the time of writing), \gls{armpl-acr} does not support the DIA format.

\subsubsection{FPGAs}

% perf advantage for small/medium matrices (see fpga_mscr)
% energy advantage over traditional architectures

Field Programmable Gate Arrays (FPGAs) provide a very large number of configurable logic components sitting within a sea of configurable interconnect. Modern FPGAs also contain hardened components, such as BlockRAM (BRAM) which provides fast on-chip memory similar to a CPU's level 1 cache and DSP slices for undertaking floating point arithmetic. Modern FPGAs are also commonly coupled with external High Bandwidth Memory (HBM2), DDR, and high performance networking capabilities. A major challenge with FPGAs has been the historically significant time investment required in programming the technology and need for detailed hardware-level knowledge on behalf of developers. Nevertheless, in recent years FPGA hardware and software development ecosystems have become far more capable, and with toolchains such as Intel's Quartus Prime and Xilinx's Vitis software developers can now program FPGAs and accelerate HPC workloads by writing code in C or \Cpp{} using High-Level Synthesis (HLS). Programming FPGAs is now becoming more a question of software development than hardware design, and consequently lowering the entry barriers for programming these devices has enabled numerous communities to investigate and explore FPGAs in their respective domains~\cite{brown2021porting,brown2021accelerating,fpga_cds}. 

Being able to tailor hardware to the code at the electronics level provides the potential to implement custom optimisation techniques around memory access and data transfer. However, to obtain best performance the programmer must rework their algorithm into a dataflow style \cite{fpga_need_for_speed} and this also often delivers much higher energy-efficiency than traditional architectures too\cite{fpga_tsukuba}. AMD Xilinx's most recent generation, known as the Versal Adaptive Compute Acceleration Platforms (ACAP)\cite{brown2023versalaie}, combines the programmable logic (PL) resources on the FPGA chip with more than 400 AI Engines (AIE). These AIE are hardened on the chip and each represent a Very Long Instruction Word (VLIW) processor capable of executing seven instructions per cycle and interconnected between AIEs and to PL on the FPGA with fast Network-on-chip (NoC). This upgrade in compute capabilities with 8-way vectorisation per AIE is especially interesting to kernels such as SpMV which exhibit large potentials in this regard. 

% In this paper we present our approach in leveraging the new AIE capabilities by implementing the common matrix storage formats COO, CSR and DIA on FPGA and incorporating our work into Morpheus. Moreover, Xilinx implemented a version of HPCG on FPGA\cite{fpga_xilinx_hpcg} and evaluating Morpheus' benefits concerning SpMV on this HPCG version completes our contribution.

% not sure how mature this is, but could look into it - again not sure at all about the required workload but happy to try

% https://github.com/Xilinx/HPCG_FPGA

% SpMM implementation on the new Xilinx Versal ACAP AI Engines, in CSR format:
%
%https://arxiv.org/pdf/2206.13734.pdf
%H-GCN: A Graph Convolutional Network Accelerator on Versal ACAP Architecture

\subsection{Sparse Matrix Storage Formats}\label{s0203}
\begin{figure}[t]
    \centering
    \captionsetup{justification=centering}
    \begin{subfigure}{0.3\columnwidth}
            \captionsetup{justification=centering}
            \includegraphics[width=\columnwidth]{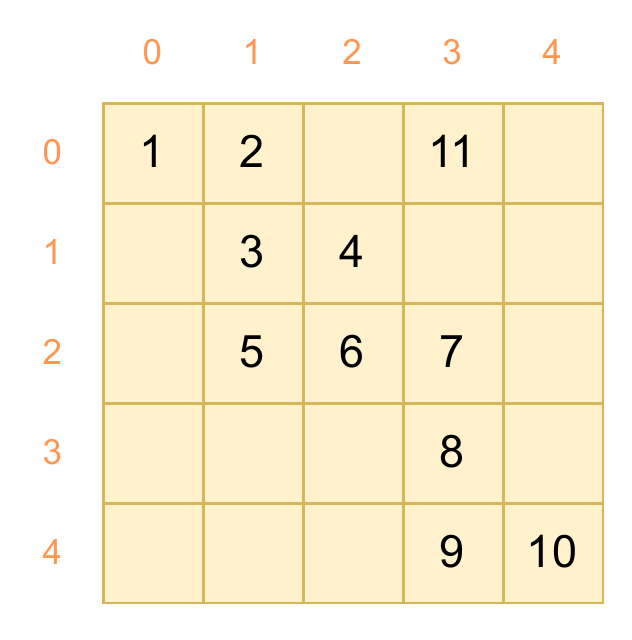}
            \caption{Dense}
            \label{fig:dense-mat}
    \end{subfigure}
    ~
    \begin{subfigure}{0.65\columnwidth}
            \captionsetup{justification=centering}
            \includegraphics[width=\columnwidth]{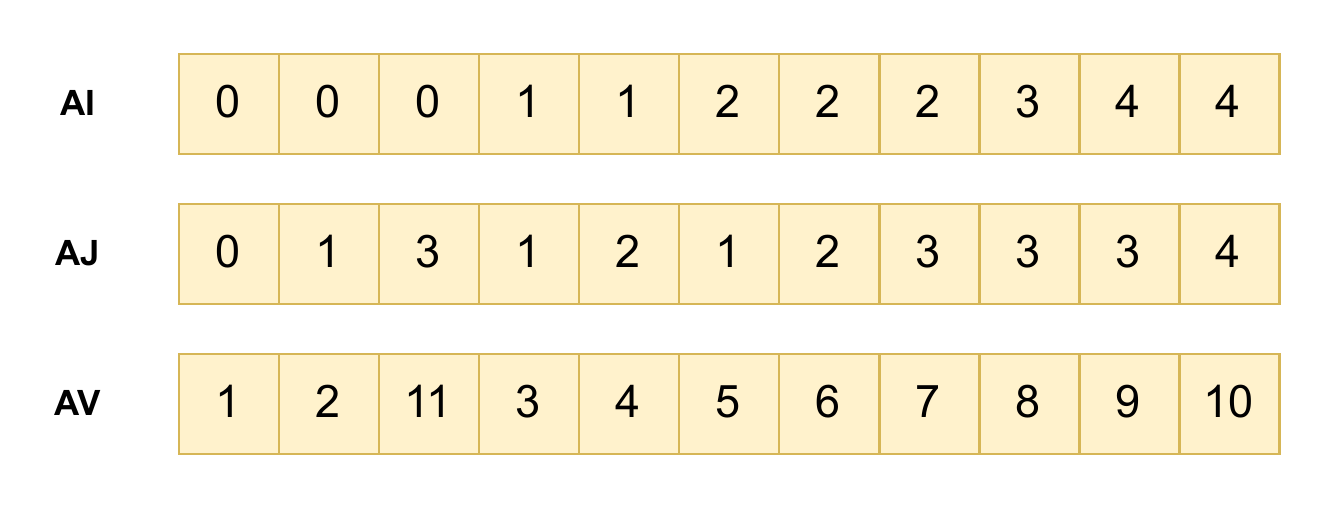}
            \caption{COO}
            \label{fig:coo-mat}
    \end{subfigure}

    \begin{subfigure}{0.65\columnwidth}
        \captionsetup{justification=centering}
        \includegraphics[width=\columnwidth]{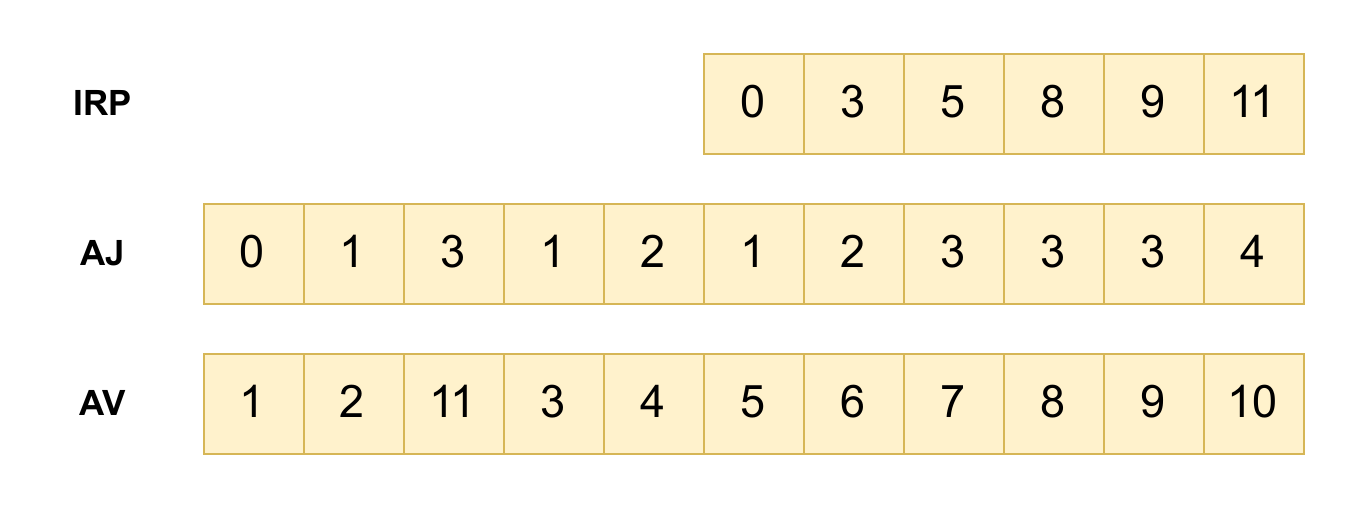}
        \caption{CSR}
        \label{fig:csr-mat}
    \end{subfigure}
    ~
     \begin{subfigure}{0.3\columnwidth}
            \captionsetup{justification=centering}
            \includegraphics[width=\columnwidth]{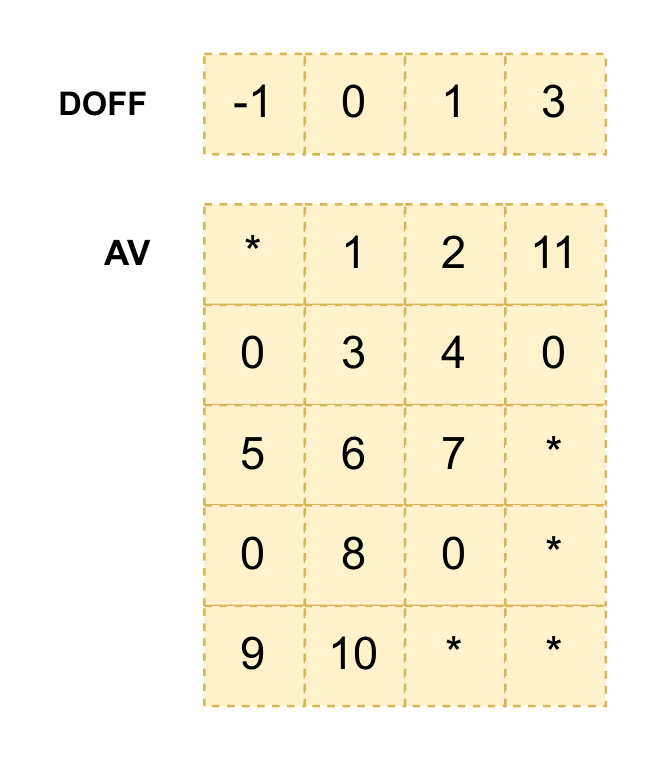}
            \caption{DIA}
            \label{fig:dia-mat}
    \end{subfigure}
    \caption{A $5\times5$ dense matrix with 11 non-zero values and its equivalent representations in 3 sparse matrix storage formats.}
    \label{fig:formats}
\end{figure}

Sparse matrices exploit the property that the majority of coefficients in the matrix are zeros by not explicitly storing those values. In other words, sparse matrix storage formats only store the non-zero coefficients and the necessary information that is required to rebuild the original position of each coefficient in the dense matrix. Each storage format rebuilds the original index of each coefficient in a different way and as a consequence each format can have very different memory layout as shown in Figure~\ref{fig:formats}. With the underlying data structure across formats varying significantly, accessing and manipulating entries in each format can result in different memory access patterns, costs and interfaces amongst formats.

Figure~\ref{fig:formats} shows the representation of a $5\times5$ dense matrix in the three sparse storage formats used throughout this work: \gls{coo}, \gls{csr} and \gls{dia}. The most basic and well-known formats are \gls{coo} and \gls{csr}. Both are considered \emph{general purpose} formats, suitable to a broad range of matrices of arbitrary sparsity patterns and target architectures. Below we provide a brief description of these formats. For a more comprehensive description please refer to Saad, Y.\cite{Saad_2003}.

\subsubsection{\gls{coo} (Figure~\ref{fig:coo-mat})} uses three arrays, whereby each non-zero element (\texttt{AV}) is explicitly stored together with its column (\texttt{AJ}) and row indices (\texttt{AI}) with no guarantees imposed in the ordering of the elements. The \gls{spmv} algorithm for \gls{coo} is shown in Algorithm~\ref{alg:coo-spmv}.

 \begin{algorithm}
 \caption{\gls{coo} \gls{spmv}}\label{alg:coo-spmv}
 \begin{algorithmic}
 \For{i=0:NNZ}
     \State y(ai(i)) += av(i)*x(aj(i));
 \EndFor
 \end{algorithmic}
 \end{algorithm}

\subsubsection{\gls{csr} (Figure~\ref{fig:csr-mat})} was implemented as an optimisation to \gls{coo}, where the \texttt{AI} array was compressed to generate an array of row pointers. As a result, \gls{csr} explicitly stores the column indices and non-zero values, and also uses an array of pointers (\texttt{IRP}) to mark the boundaries of each row, thereby reducing the memory footprint of the format by essentially compressing the row indices. As the row pointers are used to represent the position of the first non-zero element in each row, and the last entry shows the total number of non-zeros in the matrix, \gls{csr} naturally imposes an ordering across rows, though not within each row. The \gls{spmv} algorithm for \gls{csr} is shown in Algorithm~\ref{alg:csr-spmv}.

 \begin{algorithm}
 \caption{\gls{csr} \gls{spmv}}\label{alg:csr-spmv}
 \begin{algorithmic}
 \For{i=0:nrows}
     \State $sum = 0;$
     \For{j=irp(i):irp(i+1)}
         \State sum += av(j)*x(aj(j));
     \EndFor
     \State y(i) = sum;
 \EndFor
 \end{algorithmic}
 \end{algorithm}

\subsubsection{\gls{dia} (Figure~\ref{fig:dia-mat})} is a specific purpose format originally designed to perform optimally on vector architectures. It is suitable for regular sparsity patterns. \gls{dia} uses a two-dimensional array, where each column holds the coefficients of a diagonal of the matrix (\texttt{AV}), and an integer offset array (\texttt{DOFF}) keeps track of where each diagonal starts. Therefore, the \gls{dia} format is suitable for matrices with structures that dominate along the diagonals, such as banded matrices that result from discretisation methods like the \gls{fdm}. The \gls{spmv} algorithm for \gls{dia} is shown in Algorithm~\ref{alg:dia-spmv}.

 \begin{algorithm}
 \caption{\gls{dia} \gls{spmv}}\label{alg:dia-spmv}
 \begin{algorithmic}
 \For{i=0:nrows}
     \State $sum = 0;$
     \For{j=0:ndiags}
         \State k = i + doff(j)
         \If{k$\ge$0 \textbf{and} k$<$N}
             \State sum += av(i,j)*x(k);
         \EndIf
     \EndFor
     \State y(i) = sum;
 \EndFor
 \end{algorithmic}
 \end{algorithm}

% Move that in experiments
% \subsection{HPCG}
% Overview of HPCG
% How Arm HPCG differs
% How Morpheus-enabled HPCG differs

%% file: src/implementation.tex
\section{\gls{spmv} implementations on AArch64 CPUs}\label{sec:SpmvAArch64}

% rjj: this section
% rjj: related work (spmv on aarch64)
% rjj: IV and V, integrations

Optimising code for general-purpose processors such as most AArch64 CPUs typically involves one of two things: at a higher level, application programmers can choose to utilise target-specific libraries that implement core algorithms and routines efficiently for the specific targets; meanwhile, at a lower level, application programmers can write efficient code targeting specific CPU (micro-)architectures themselves, usually either through intrinsics (or built-ins), which are functions treated specially by compilers to make features of the target architecture directly available to programmers, or via writing explicit assembly for their target. In this work we have explored integrating these two forms of optimisations for AArch64 CPUs into \emph{Morpheus}, which we describe in this section.

The Arm Performance Libraries (ArmPL)~\cite{armpl} are a set of core routines developed by Arm for HPC applications
for AArch64 targets (especially Neoverse-based). It contains BLAS, LAPACK, FFT, Sparse,
libamath (a subset of libm) and libastring (a subset of libc for strings) routines
for both single- and multi-threaded processing provided via both C and Fortran interfaces.
ArmPL's sparse routines support dense, CSR, CSC, COO and BSR matrices.
These routines are provided via an API similar to FFTW, where the description of the problem is independent from its execution. In this sense, to set up an SpMV operation with ArmPL we start by creating a handle to a sparse matrix. This is achieved with the \verb|armpl_spmat_create_*| family of routines. Usually, this matrix is provided in a common format such as CSR. In our case, however, the handle is created for the specific format we are using at the time. Then, hints are provided to the handle in an attempt to speedup future SpMV calls with \verb|armpl_spmat_hint| calls. The handle is then used in an optimisation stage similar to that found in other libraries such as the aforementioned FFTW, where the library tries to determine the best algorithms and implementations for the specific matrix and target. This step is issued with \verb|armpl_spmv_optimize|. Once these optimisations have been run, the handle can be used repeatedly to execute SpMV and other sparse algebra computations via the \verb|armpl_*_exec_*| family of routines. Once the handle is not needed anymore, it can be destroyed by calling \verb|armpl_spmat_destroy|. Given ArmPL's high-performance for AArch64 targets and ease-of-use, we have chosen it in this paper to explore how and how well target-specific libraries (of which ArmPL is an example) can be integrated into \emph{Morpheus}, thus offering its high-performance to \emph{Morpheus} users transparently.

An alternative way of developing highly-optimised code for AArch64 CPUs is by leveraging the Arm C Language Extensions (ACLE)~\cite{acle}. The ACLE are a set of compiler intrinsics that expose advanced features of the Arm achitecture and aim to enable the development of applications and libraries portable across compilers and across Arm micro-architectures. One of the most disruptive extensions of the Arm architecture (in particular AArch64) is the Scalable Vector Extension (SVE)~\cite{sve}. Unlike other contemporary single instruction multiple data (SIMD) extensions such as Neon from Arm and the AVX extensions from x86, SVE is a ``vector-length-agnostic'' (VLA) vector extension. This means that the programmer does not program to vector registers of specific width; instead, they program to a slightly different programming model where the width of the vector registers is not known at compile-time. In practice, this fact makes SVE highly portable, as the same code (and, in fact, binary) can run transparently on hardware with vector registers of different widths. Besides offering this extra flexibility and portability, SVE is also an extremely complete instruction set, supporting instructions highly suitable for High Performance Computing (HPC) and Machine Learning (ML) applications. Some examples of the high-performance features of SVE are per-lane predication (i.e. control on a per vector element basis), gather-loads and scatter-stores, speculative vectorisation, and horizontal and tree-based reductions. The gather-loads/scatter-stores of SVE are especially useful for SpMV computations given the latter's irregular and indirect access patterns (as exemplified in Section~\ref{s0203}). Due to these reasons, in this work we used ACLE to develop SVE-enabled implementations of SpMV kernels for COO, CSR and DIA matrices. Our implementations result mostly from a transliteration of the default \Cpp versions of the SpMV kernels present in \emph{Morpheus} to ACLE. Below we highlight two of the main implementation details of our SVE-enabled SpMV kernels. We utilise the algorithms presented in Section~\ref{s0203} throughout to facilitate the discussion.

The indirection in the output vector \texttt{y} through the row index vector \texttt{ai} in the COO kernel complicates the vectorisation of the kernel's loop. This happens because independent elements of \texttt{ai} might point to the same element of \texttt{y}. In these cases, the writes to \texttt{y} have either to be serialised or accumulated before being issued, so that a single write is effected and no updates to \texttt{y} are lost. In our SVE-enabled COO implementation we have chosen the second approach, whereby in each iteration in \texttt{i} we only work with the elements \texttt{(ai(i), ai(i+1), ai(i+2), ...)} that match (i.e. are equal to) \texttt{ai(i)}. 
We leverage SVE's predication features to create a mask of such elements and only operate on them.
This allows us to accumulate the products \texttt{(av(i)*x(aj(i)), av(i+1)*x(aj(i+1)), ...)} that correspond to the same \texttt{ai(i)} before writing them to \texttt{y}, thereby effecting a single accumulation in \texttt{y}. In C and ACLE pseudocode, this correspond to:
\begin{minted}[fontsize=\footnotesize, linenos, xleftmargin=1em, numbersep=5pt]{c}
vbool_t pg;
for(i = 0; i < NNZ; i += vcntp(pg, pg)) {
  // Generate mask for the values of i < NNZ
  pg = vwhilelt(i, NNZ);
  
  // Load values ai(i, i+1, ...)
  vidx_t vai = vld1su(pg, ai+i);
  
  // Generate mask for the elements
  // (ai(i), ai(i+1), ...) == ai(i)
  pg = svcmpeq(pg, vai, ai[i]);

  // Load values of aj, av, and x
  vidx_t vaj = vld1su(pg, aj+i);
  vtype_t vav = svld1(pg, av+i);
  vtype_t vx = svld1_gather_index(pg, x, vaj);

  // Compute products av(i)*x(aj(i))
  vtype_t vr = svmul_x(pg, vav, vx);
  
  // Accumulate products
  yval[ai[i]] += svaddv(pg, vr);
}
\end{minted}
Though this approach might be inefficient if the mask \texttt{pg} becomes too ``hollow'' (i.e. with too few active elements), in practice in our tests this does not happen frequently. Thus, as shown in Figure~\ref{fig:arm-spmv}-a, this strategy allows us to achieve significant speedups over default (i.e. compiler generated) and ArmPL implementations.

Another important subtlety in the way we implement our SVE-enabled SpMV kernels lies in the way we vectorise the DIA format. Instead of vectorising the inner loop (in \texttt{j}), as is more common, we have vectorised the outer loop (in \texttt{i}). We have done this for two reasons, namely (i) the memory accesses in \texttt{av} are contiguous (i.e. with stride 1) in \texttt{j}, thus we get better cache utilisation from loading several \texttt{(av(i, j), av(i+1, j), ...)} at a time and then looping through \texttt{(j, j+1, ...)} sequentially, and (ii) this avoids doing a horizontal reduction to accumulate the values of \texttt{sum} before writing it to \texttt{y}. Additionally, we once again resort to SVE's predication features to mask the valid \texttt{k} indices for each inner iteration. In pseudocode, we have:
\begin{minted}[fontsize=\footnotesize, linenos, xleftmargin=1em, numbersep=5pt]{c}
vidx_t vidx = vindex(0, ndiags);
for(i = 0; i < rows; i += vcnt()) {
  // Initialise sum
  vtype_t vsum = vdup(0);
  
  // Create mask for the values of i < nrows
  vbool_t pg = vwhilelt(i, nrows);

  for(index_type j = 0; j < ndiags; j++) {
    index_type k = i + doff[j];

    // Generate mask for the valid k's
    // p1 = k < 0
    // p2 = k < N
    // pm = p2 && !p1
    vbool_t p1 = vwhilelt(k, 0);  
    vbool_t p2 = vwhilelt(k, N);  
    vbool_t pm = svbic_z(pg, p2, p1);

    // Load av and x
    vtype_t vav =
      svld1_gather_index(pm, av+i*ndiags+j, vidx);
    vtype_t vx = svld1(pm, x+k);
    
    // Compute the products av(i, j)*x(k) and
    // accumulate them
    vsum = svmla_m(pm, vsum, vav, vx);
  }

  // Store the results in (y(i), y(i+1), ...)
  svst1(pg, y+i, vsum);
}
\end{minted}
Once again, it might happen that the predicate \texttt{pm} might have too few elements active for vectorisation to pay off, though this does not happen often. Furthermore, in our tests (Figure~\ref{fig:arm-spmv}-c) the choice of performing outer-loop vectorisation over inner-loop vectorisation leads to significant speedups. However, we note that the compilers we tested, namely GCC 11.2.0 and LLVM 15.0.7, are not able to perform this outer-loop vectorisation automatically due to the complex control flow it entails.

% food for thought: we can brand Morpheus as an aggregator of SpMV storage formats and algorithms

\section{\gls{spmv} implementations on \glspl{fpga}}\label{sec:spmvFPGA}
\label{sec:fpga_limitations}

%\textcolor{red}{Gaining performance on FPGA: Howto}
\glspl{fpga} provide programmable logic that can be configured at the electronics level to represent the hardware tailored to a specific algorithm. The way that \glspl{fpga} operate fundamentally differs from how \textit{Von-Neumann} architectures such as traditional CPUs operate, therefore during porting of CPU-based algorithms these need to be re-engineered to a dataflow style of computing suitable for such devices\cite{fpga_need_for_speed}. \textit{Dataflow}, a fundamental concept for the quest of performance on FPGAs, is typically built around concurrently running stages, known as dataflow stages, that stream data between themselves and each stage comprises individual pipeline(s) which will start to process a new iteration each cycle. %With algorithms of high throughput requirements, pipeline stalls in FPGA programming can often be observed in scenarios with loop-carried dependencies [29] and with resource contention when memory interfaces face concurrent accesses [27]. Mitigating these overheads, HLS toolings provide approaches such as array partitioning to enable parallel access of memory locations across multiple memory ports or banks to reduce contention and ultimately improve loop throughput.
On AMD-Xilinx \glspl{fpga}, which are the focus of this work, this regularly includes the definition of \textit{dataflow regions} which are connected through HLS streams providing usually lower number of cycles from reads and writes than accesses to global memory.

% the HLS tooling completely unrolls \texttt{acc_partial_loop} and pipelines the \texttt{nnnz_loop} with II=8, essentially generating a result every clock cycle

Porting the three \emph{Morpheus} storage formats COO, CSR and DIA to the FPGA, we implement each of these as individual kernels and deliver three different bitstreams, which configure the FPGA, each containing a kernel of the SpMV of the respective storage format algorithm as presented in Algorithm~\ref{alg:coo-spmv} for COO, Algorithm~\ref{alg:csr-spmv} for CSR and Algorithm~\ref{alg:dia-spmv} for DIA. Working with AMD-Xilinx's HLS toolchain Vitis in C/C++, this means we had to set up the host code on CPU and the HLS kernels on the AMD-Xilinx Alveo U280 FPGA which is used in this work. AMD-Xilinx's host-device model is built upon OpenCL, and as such we (i) initialise the device in the host code, (ii) create OpenCL buffers for input and output data, (iii) transfer the required input matrix and vector data to the device, (iv) execute the kernel on device once data has been transferred and is available in the global device memory and lastly (v) transfer the result vector back to the host. Depending on the matrix dimensions, the required input and output data size varies and we transfer data and run the device kernels in single precision floating-point. While the number of elements and data sizes of input data for COO and CSR are relatively similar, the DIA format requires substantially more matrix values depending on the number of padded rows and the number of diagonals to be stored: for COO all inputs (\texttt{AI}, \texttt{AJ}, \texttt{AV}) require \texttt{n=NNZ} elements with data sizes of 32-bit for integers and 32-bit for floats, for CSR \texttt{IRP} requires \texttt{n=nrows$+$1} elements of 32-bit for integers, \texttt{AJ} and \texttt{AV} both require \texttt{n=NNZ} elements of 32-bit for floats, and for DIA \texttt{DOFF} requires \texttt{n=ndiags} for 32-bit integers and \texttt{AV} requires \texttt{n=padded\_rows*ndiags} elements of 32-bit for floats. Compared to the input data requirements of COO and CSR, the DIA format requires significantly more input elements and therefore increases the amount of data to be transferred from host to device.

Under the assumption that the majority of SpMV computations occur as one out of many routines within a larger kernel such as in HPCG, it is reasonable to assume the availability of input data on the device and therefore we only report kernel execution time on the device (excluding device initialisation and setup time to download the bitstream and configure the hardware and excluding data copy on and data copy off time). Our implementation provides the data copy to device and data copy back to host each in a single OpenCL buffer and with the Vitis toolchain this means that we are bound to the tooling's single buffer size limit of 4~GB\footnote{https://docs.xilinx.com/r/en-US/ug1393-vitis-application-acceleration/Buffer-Creation-and-Data-Transfer}, inhibiting us from running matrices that require individual input buffers larger than the buffer size limit. For the Alveo U280 FPGA, there is 8~GB of High-Bandwidth Memory (HBM) and 32~GB of DDR DRAM available on-card. We utilise the faster HBM for data transfers to device, which is why we do not run larger matrix benchmarks than those requiring less than 8~GB of accumulated input data for the respective SpMV algorithms.

%\textcolor{red}{Important: A single buffer cannot be bigger than 4 GB, yet to maximise throughput from the host to global memory, Xilinx also recommends keeping the buffer size at least 2 MB if possible.}
%\subsection{Optimisations}

\begin{figure}[t]
    \centering
    \includegraphics[width=0.9\columnwidth]{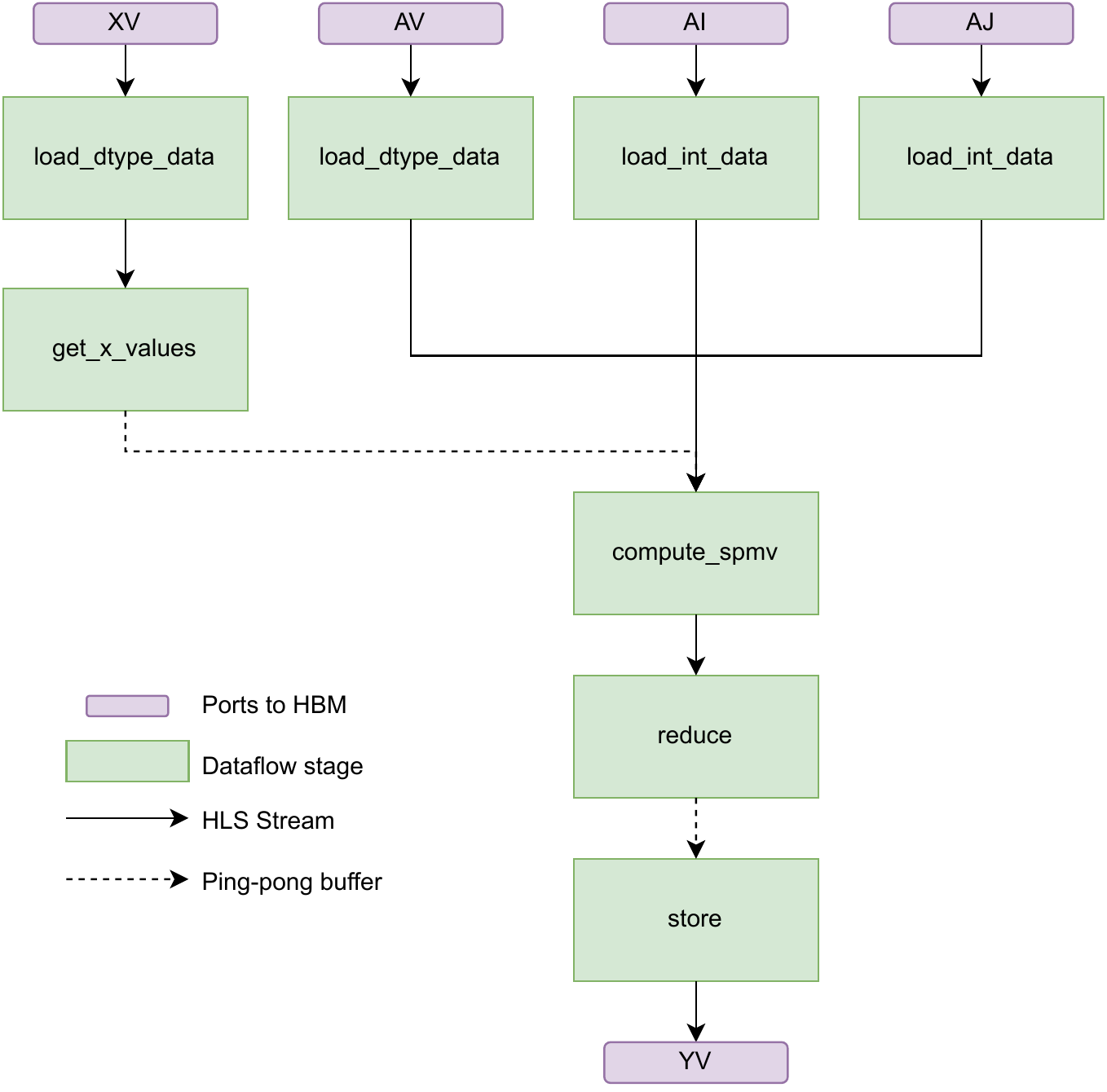}
    \setlength{\belowcaptionskip}{-8pt} 
    \caption{Dataflow region of the COO algorithm implementation on FPGA with dataflow stages operating in parallel and connected through HLS streams and ping-pong buffers implemented in UltraRAM.}
    \label{fig:fpga_coo_dataflow_region}
\end{figure}

Figure~\ref{fig:fpga_coo_dataflow_region} illustrates the structure of our dataflow version of COO on the FPGA using AMD Xilinx Vitis HLS. Each green box is a separate dataflow region running concurrently, with arrows between these illustrating streams of data that flow from one cycle to the next. The purple boxes in Figure~\ref{fig:fpga_coo_dataflow_region} depict connection to external, high bandwidth, memory where all reads and writes are packed in chunks of 512 bits to best utilise the memory controllers. Dataflow regions run in parallel to load the data from HBM2 and then pass individual data elements to the next stages. The dashed line in Figure~\ref{fig:fpga_coo_dataflow_region} represents a ping-pong buffer, which is a common double buffering technique used in FPGA programming where the dataflow stage will concurrently write to one buffer whereas the subsequent stage is served with data from a previous copy of the buffer and these then switch at a predefined point.

One area of concern is to ensure that pipelines that comprise the dataflow regions have initiation intervals (II) of one, where the II refers to the number of clock cycles before the next iteration in a loop can be started. An initiation interval of one, commonly written as \texttt{II=1}, means that the dataflow design can essentially yield a result every cycle. Considering the much slower clock frequency of FPGAs, typically around 300MHz, compared to CPUs and GPUs, for performance it is critically important that every cycle counts when it comes to computation. %On the Alveo U280, there is UltraRAM (URAM) with 960 RAMs of 288KB each available, which we fill up with three \texttt{float32} buffers for \texttt{XV}, \texttt{AV} and \texttt{YV}, each of 23,040,000 elements at maximum (based on the \texttt{MAX\_NROWS} and \texttt{MAX\_NCOLS} constants).  

The following pseudocode describes our \texttt{reduce} stage in the dataflow region of Figure \ref{fig:fpga_coo_dataflow_region} which illustrates our approach in moving to an optimal II of 1:

\begin{minted}[fontsize=\footnotesize, linenos, xleftmargin=1em, numbersep=5pt]{c}
void reduce(const unsigned int A_nrows, 
  const unsigned int A_nnnz, 
  unsigned int A_rind[MAX_NROWS], 
  hls::stream<dtype> &sum_stream, 
  dtype y_val[MAX_NROWS]
) {
nrows_loop:
for (unsigned int row_index=0; row_index<A_nrows; 
  row_index++) {
  dtype acc_part[LATENCY]={0,0,0,0,0,0,0,0};
nnnz_loop:
  for (unsigned int i=0; i<A_nnnz; 
    i+=LATENCY) {
    #pragma HLS pipeline
acc_partial_loop:
    for (unsigned int j=0; j<LATENCY; j++) {
      #pragma HLS unroll
      dtype sum = sum_stream.read();
      unsigned int tmp_A_rind = 
        A_rind[i*LATENCY+j];
      if(tmp_A_rind == row_index) {
        acc_part[j] += sum;
      }
    }
  }
acc_final_loop:
  for (unsigned int j=0; j<LATENCY; j++) { 
    #pragma HLS unroll 
    y_val[row_index] = acc_part[j];
  }
}}
\end{minted}

% Naive COO:
%for (int n = 0; n < A_nnnz; n++) {
%    y_val[A_rind[n]] += A_val[n] * x_val[A_cind[n]];
%}

% DISCLAIMER: looking at the performance results from section "evaluationn", we can say that adding the outermost loop over A_nrows we can do the matching for row ids (conditional comparing row ids), and these are perfectly pipelined II=1 but compared to the naive problem we waste too many cycles doing nothing (ie. the conditional is true not very often)

Implementing the naive reduction to \texttt{YV}, as part of the COO Algorithm~\ref{alg:coo-spmv}, based on the \textit{row index} of each non-zero element on FPGA as \texttt{fadd} required eight clock cycles. The HLS tooling detected a spatial dependency and therefore avoided pipelining our \textit{reduce} function at lower II than eight. Splitting the reduction into \texttt{LATENCY=8} partial accumulations in the inner loop \texttt{acc\_partial\_loop} on Line~15 in the pseudocode above, which the tooling can fully unroll, the outer \texttt{nnnz\_loop} will still be pipelined with II=8 by the HLS tooling but through the full inner unroll essentially yields a result every cycle. As in the \gls{coo} algorithm we only reduce across the same \textit{row index}, on Line~21 we introduce a conditional to only accumulate the partial sum for the same \textit{row index}. On Line~26, we then perform the final accumulation across the previously computed partial accumulations and write the reduced value to \texttt{YV} (here \texttt{y\_val}) based on the corresponding \textit{row index}. The presented reduction builds on a \texttt{LATENCY} of eight for the \texttt{fadd} which required eight cycles, and to ensure that none of the HLS streams contain leftover data at the end of the SpMV kernel, on the host before creating OpenCL buffers and transferring data to the FPGA's global memory, we apply padding to all input data structures to be multiples of \texttt{LATENCY} (here eight). With our approach, the optimised SpMV kernel is not bound to any specific matrix dimensions, for instance does also run on matrices with row and column numbers which are not multiples of eight but is still limited by individual buffer size and accumulated input buffer sizes.% as explained in section \ref{sec:fpga_size_limits}. 

% optimised COO on FPGA with local buffers such as \texttt{A_rind[MAX_NROWS]} implemented in URAM which is
% URAM is 288KB * 960 rams = 276,480 KB (or ca 270 MB)
% divided by three buffers = 92,160 KB per buffer
% in float32 means 23,040,000 elements per buffer
% (three buffers are: x_val_buf, y_val_buf, A_rind_buf
% const unsigned int MAX_NCOLS=23040000;
% const unsigned int MAX_NROWS=23040000;

% reduce / accumulate stage is as follows:

% LATENCY is 8 for fadd with select
% on host, we add padding to all involved arrays/inputs to avoid leftover data
% therefore on device, we can always reduce with interleaving factor=LATENCY

\section{Integration with Morpheus}
\emph{Morpheus} is a \Cpp header-only library that heavily relies on templates and meta-programming to enable certain polymorphic capabilities. It follows a functional design that separates the data structures (containers) from the functions (algorithms), with algorithms acting on containers. In order to provide support for the various hardware platforms and memory hierarchies, \emph{Morpheus} adopts two notions of abstraction: 1)~the \emph{Execution Space}, which specifies where the code will be executed; and 2)~the \emph{Memory Space}, specifying where the data will \textit{reside in memory}. \emph{Morpheus} currently supports four executions spaces: 1)~Serial (Sequential), 2)~OpenMP (Multi-threaded), 3)~CUDA (NVIDIA GPUs) and 4)~HIP (AMD GPUs), with each execution space also acting as a separate backend. As a result, it is possible to use a single interface for each supported algorithm. By specifying the backend we want to run in, we also target a different execution space. In addition, \emph{Morpheus} offers data management routines to effectively managing data transfers between the available memory spaces. To provide support for heterogeneous hardware, \emph{Morpheus} adopts the \emph{Host-Device} model, enabling data management functionality between different memory spaces.

The integration of specific optimisations, such as the ones proposed in Section~\ref{sec:SpmvAArch64} in a pre-existing backend presents different challenges that needed to be overcome from the integration of a new backend as proposed in Section~\ref{sec:spmvFPGA}. Below we provide a description of the challenges for each of the two approaches.

\subsection{Optimisations}
\emph{Morpheus} performs compile-time introspection on the algorithm and by examining the backend, provided by the user as a parameter, identifies which algorithm implementation to dispatch every time. This means it is only possible to dispatch one and only implementation of the algorithm for every backend. Multiple implementations in a single backend can be chosen only by re-compiling the code.

The optimisations proposed at Section~\ref{sec:SpmvAArch64} effectively constitute alternative algorithm implementations of the \gls{spmv} routine in the Serial backend. With the current state of \emph{Morpheus}, the integration is only possible by specifying which version should be enabled using a compile-time flag. Of course future developments could move this decision at run-time, such that it will be possible to enable all versions, although this will introduce runtime overheads every time the algorithm is executed. However, this would only be possible if the performance improvements of selecting the most optimal version (given the sparsity pattern of the matrix) justify this trade-off.

The integration of the \gls{spmv} implementations for \gls{coo} and \gls{csr} offered by \gls{armpl-acr} inside each of the equivalent \emph{Morpheus} routines requires the data of the \emph{CooMatrix} and \emph{CsrMatrix} containers in \emph{Morpheus} to be converted in \emph{armpl\_spmat\_t} so that they will be passed to the \gls{armpl-acr} specific routines. Since the data layout of the \emph{CooMatrix} and \emph{CsrMatrix} follows the same requirements as the ones internally in \gls{armpl-acr}, the \emph{armpl\_spmat\_t} handle is created in \texttt{ARMPL\_SPARSE\_CREATE\_NOCOPY} mode, meaning that only the pointers to the data are passed instead of any actual copies. To avoid creating a new handle for the same matrix every time the \gls{spmv} multiplication is performed one workspace is created for each format acting as a Singleton~\cite{design_patterns}. In other words, every time the \gls{spmv} multiplication is executed with a new matrix the workspace is responsible to register the newly created handle and in future calls use that one instead of creating a new one. Another challenge that had to be resolved is the adaptation of the polymorphic behaviour of \emph{Morpheus} containers to explicit \gls{armpl-acr} function calls that have the format and value type information embedded in the function name. An example of the adaptor call that had to be implemented that maps the polymorphic containers available in \emph{Morpheus} to the explicit function call for creating the \gls{armpl-acr} handler of either a \gls{coo} or \gls{csr} matrix using \gls{armpl-acr} is shown in Table~\ref{tab:polymoprhic-mapping}.

\begin{table}[h]
    \caption{Adaptation of the polymorphic \emph{Morpheus} behavior to explicit calls required by \gls{armpl-acr}.}
    \label{tab:polymoprhic-mapping}
    \begin{center}
    \begin{small}
    \resizebox{\columnwidth}{!}{%
    \begin{tabular}{c|c|c}
        Morpheus Format &  Adaptor call & \gls{armpl-acr} call\\
        \hline
        \texttt{CooMatrix$<$double$>$} & \multirow{2}{*}{\texttt{create\_coo$<$T$>$}} & \texttt{armpl\_spmat\_create\_coo\_d}\\
        \texttt{CooMatrix$<$float$>$}  & & \texttt{armpl\_spmat\_create\_coo\_s}\\
        \hline
        \texttt{CsrMatrix$<$double$>$} & \multirow{2}{*}{\texttt{create\_csr$<$T$>$}} & \texttt{armpl\_spmat\_create\_csr\_d}\\
        \texttt{CsrMatrix$<$float$>$}  & & \texttt{armpl\_spmat\_create\_csr\_s}\\
    \end{tabular}}
    \end{small}
    \end{center}
\end{table}

After a successful porting of the \gls{armpl-acr} implementations in \emph{Morpheus}, the port of the individual \gls{sve} implementations for each format as described in Section~\ref{sec:SpmvAArch64} followed the same principles as described for \gls{armpl-acr}. In other words, since no handles were used in the \gls{sve} implementations, no workspaces were required. However, the \gls{sve} intrinsics used in the implementations had to be adapted in a similar way described for \gls{armpl-acr}.

\subsection{New Backends}
The integration of a new backend in any code is a very challenging task as each new backend poses unique challenges that might require significant development efforts. In this work, we seek to understand the challenges involved in enabling support for \glspl{fpga} in \emph{Morpheus} for future releases. The motivation behind this choice is that \glspl{fpga} share many similarities with GPUs because, from the developers perspective, both are regarded as accelerators with distinct memory spaces from the host (CPU) in a \textit{host-device model}. Hence, since \emph{Morpheus} already provides supports for GPUs, the high-level interface for managing heterogeneous hardware could potentially remain unchanged.

The proposed process of support for \gls{fpga} in \emph{Morpheus} is as follows: 
\begin{enumerate}
    \item \textit{Develop the execution space:} In order to maintain portability of \emph{Morpheus}, the \emph{ExecutionSpace} concept available in \emph{Kokkos}~\cite{kokkos} can be followed in order to implement a common set of functionality that remains in line with the interface of the existing execution spaces. Upon completion, \emph{Morpheus} will be able to discover, acquire and synchronise with an \gls{fpga} that is available in the runtime and dispatch any algorithms implemented for the \gls{fpga} backend. Note that housekeeping routines from OpenCL might be required for querying the underlying \gls{fpga} runtime.
    \item \textit{Develop the memory space:} In a similar manner to the creation of the execution space, the memory space for the global memories that are available on \glspl{fpga} has to be developed. Following the \emph{MemorySpace} concept available in \emph{Kokkos}, two memory spaces can be created representing the DDR and \gls{hbm}, memories that are commonly available in many \glspl{fpga}. Upon creation, using the two memory spaces users will be able to allocate/deallocate memory on the target \glspl{fpga} effectively creating any of the available containers on the FPGA.
    \item \textit{Data Management:} Following the development of memory spaces that can obtain memory on the FPGA, the data management routines for each of the new memory space must be implemented. These routines include \emph{copy} and \emph{mirroring} operations responsible for transferring data between the supported memory spaces and creating new containers with the same characteristics and memory allocation size in a specified memory space. In addition, this will allow for data to be offloaded from the Host (CPU) to the Device (FPGA) and vice-versa.
    \item \textit{Low-level Implementation:} The last step will be to implement the algorithms available in \emph{Morpheus} (including \gls{spmv} multiplication) for the FPGA. The implementations at this level will be the ones launched when the \gls{fpga} execution space is selected. Note that both the housekeeping done with OpenCL as well as the kernel launch are implemented at this level.
\end{enumerate}

The first three steps described above share many similarities with the existing backends in \emph{Morpheus}. However, the most challenging stage to implement is the low-level implementation as this is where the unique properties of \glspl{fpga} are manifested. One major challenge is the vast difference in the build process of the device code. Generally, the code written for \glspl{fpga} is compiled to generate a bitstream containing the hardware configuration, which even for small kernels such as \gls{spmv} can take several hours, especially with increasing solution spaces for optimisations such as the partitioning of large arrays. Compared to other architectures, this reconfigurability comes at the expense of relatively long build processes including routing and placing on the die for \glspl{fpga} leading to longer time-to-solution. While building the host code on CPU for managing the interaction with the device is relatively fast compared to the overall bitstream generation, this means that on-the-fly compilation and linking is not possible. In addition, \emph{Morpheus} is a header-only library that is effectively compiled at the application level. Consequently, all the algorithms are written in the form of templates that the compiler is responsible for generating at the application level i.e. in user's code. As a result, the type of the inputs is generic until last minute meaning we wouldn't been able to generate the \gls{fpga} bitsream until that point. To circumvent this issue, a set of potential types could be used to explicitly instantiate these implementations during the build and installation of \emph{Morpheus}, such that it will be possible to generate the bitstreams \emph{apriori}.

Another important challenge is portability across different \gls{fpga} devices both of the same or different vendors. Klaisoongnoen et al.~\cite{fpga_option_streaming} explored the porting of HLS kernels between AMD-Xilinx and Intel \glspl{fpga} and described the challenges encountered when moving from one \gls{fpga} architecture to another and suitability of optimisation techniques between vendor tool chains. For the \gls{spmv} kernels in HLS presented in this work, moving from one \gls{fpga} architecture to another requires that connectivity configurations, for instance mapping kernel arguments to memory spaces, are adapted. An important consideration is whether the target \gls{fpga} card provides \gls{hbm} memory or whether alternatively on-card DDR has to be targeted and how this is integrated in the specific build process of the available HLS tooling. While the HLS code tends to be portable between \gls{fpga} architectures, vendor specific \texttt{pragmas}, host code implementations and support for features such as \textit{direct host-kernel streaming}\footnote{Intel \gls{fpga} Programming Guide, Direct Communication with Kernels via Host Pipes: \url{https://www.intel.com/content/www/us/en/docs/programmable/683846/22-1/direct-communication-with-kernels-via.html}} vary. One possible solution to this issue is the creation of different backends for each vendor as well as implementing the algorithms and optimisations with backward compatibility in mind. Note that by explicitly instantiating the \gls{fpga} implementations as well as having information about the range of the supported \gls{fpga} devices can allow us to generate a set of bitstreams in advance and distribute them as part of the \emph{Morpheus} release and each time load the appropriate pre-built implementation depending on the respective runtime.

Focusing on automatic code generation for multiple backends, a key challenge with FPGAs is the fundamental difference in how such devices operate compared to traditional \textit{Von-Neumann} based CPU architectures. Whilst HLS toolchains such as AMD-Xilinx's Vitis HLS and Intel's Quartus Prime Pro typically generate working bitstreams for FPGAs from CPU-based codes, these bitstreams tend to be significantly slower in performance compared to manually tuned HLS code which has been adapted to suit a dataflow style of computing optimal for FPGAs. For instance, differences in runtime performance between naive CPU-based HLS kernels and \textit{dataflow-optimised} implementations have been shown to range significantly (over 1000 times is common)\cite{fpga_data_movement}. Whilst available HLS tools provide portability within limitations, performance portability between CPU and FPGA architectures remains an open challenge. Our approach, as described above, is not based on automatic code generation but instead allows developers to write custom code for \glspl{fpga} with explicit optimisations. In addition, users can exploit additional optimisations through the dynamic format switching capabilities of \emph{Morpheus} that would also be available for the \gls{fpga} backend.

%% file: src/experiments.tex
\section{Results and Evaluation}
\subsection{Setup}\label{sec:setup}
All experiments targeting AArch64 CPUs were carried out on the Bristol-based HPE Apollo 80 partition of Isambard\cite{isambard}. Each of the 72 nodes on the cabinet has a Fujitsu A64FX Processor with 48 ARMv8.2 cores and 512-bit \gls{sve}, running at the clock frequency of 1.8GHz, and 32GB \gls{hbm} memory arranged in 4 core memory groups.

Each experiment was compiled with \emph{GNU 10.2.0} using \texttt{-O3} \texttt{-ffast-math} \texttt{-ftree-vectorize} \texttt{-funroll-loops} \texttt{-mcpu=native} compiler flags. For the distributed experiments \emph{OpenMPI 4.1.0} was also used.

%All CPU runs are undertaken by threading using OpenMP across two 24-core Xeon Platinum (Cascade Lake) 8260M CPUs at 2.40GHz which are fitted into a single node of our test system and energy measured via RAPL\cite{rapl}. All CPU runs are executed across all 48 physical cores as this was found to be the optimal single-node CPU configuration. 
For the FPGA runs reported in this paper we use a Xilinx Alveo U280, running at the default clock frequency of 300MHz, which contains an FPGA chip with 1.08 million LUTs, 4.5MB of on-chip BRAM, 30MB of on-chip UltraRAM, and 9024 DSP slices. This PCIe card also contains 8GB of \gls{hbm} and 32GB of DDR DRAM on the board. %We also use a Bittware 520N-MX which contains an Intel Stratix-10 MX2100 FPGA, with 702720 ALMs, 29.9MB of on-chip memory, and 3960 DSP blocks. This FPGA also contains 16GB of HBM2 external memory. The clock on the Intel Stratix-10 is more dynamic, achieving 320MHz with one kernel but this then decreases to 254MHz as we scale.

The FPGA card is hosted in the ExCALIBUR H\&ES FPGA testbed\footnote{ExCALIBUR H\&ES FPGA testbed, Field Programmable Gate Arrays (FPGAs) for accelerating scientific and data-science codes: \url{https://fpga.epcc.ed.ac.uk/}} system with a 32-core AMD EPYC 7502 CPU with 256GB DRAM %and energy metrics gathered via the \gls{xrt} and AOCL respectively
. All bitstreams are built for the U280 using Xilinx's Vitis framework version 2021.2.% which at the time of writing is the latest compatible version for the U280\%.%, and Quartus Prime Pro 20.4 for the Stratix-10. 
 All reported results are averaged over ten runs and FPGA run-time includes on-device execution time exclusive device setup time and excluding data transfer times.% and any required data reordering on the host. %Configurations and environments are described in Appendix \ref{sec:ad}.

\subsection{Evaluation of \gls{spmv} on AArch64 CPUs}\label{sec:arm-spmv}
In order to evaluate the performance of the newly added \gls{spmv} implementations for AArch64 CPUs in \emph{Morpheus}, for each implementation we perform 100 iterations of the \gls{spmv} multiplication over 2106 sparse matrices available in SuiteSparse\cite{suitesparse} collection. Each run is executed in Serial on a Fujitsu A64FX Processor, as described in Section~\ref{sec:setup}. The implementations are divided in three versions as shown in Table~\ref{tab:arm-implementations}, along with a short description and the supported formats for each. 
\begin{table}[h]
    \centering
    \caption{Versions of each CPU-based \gls{spmv} implementation available in Morpheus along with the formats each version supports.}
    \label{tab:arm-implementations}
    \begin{tabular}{c|c|c|c|c}
    Version     & Description & \gls{coo} & \gls{csr} & \gls{dia} \\
    \hline
    \multirow{2}{*}{Plain}     &  Original implementations & \multirow{2}{*}{$\checkmark$} & \multirow{2}{*}{$\checkmark$} & \multirow{2}{*}{$\checkmark$} \\
         &  without any Arm Optimisations &  &  &  \\
    \hline 
    ARMPL     & Implementations using ArmPL & $\checkmark$ & $\checkmark$ & $\times$ \\
    \hline
    \multirow{2}{*}{SVE}       & Implementations using & \multirow{2}{*}{$\checkmark$} & \multirow{2}{*}{$\checkmark$} & \multirow{2}{*}{$\checkmark$} \\
           & SVE Extensions &  &  &  \\
    \end{tabular}
\end{table}

\begin{figure}[t]
    \centering
    \includegraphics[width=0.8\columnwidth]{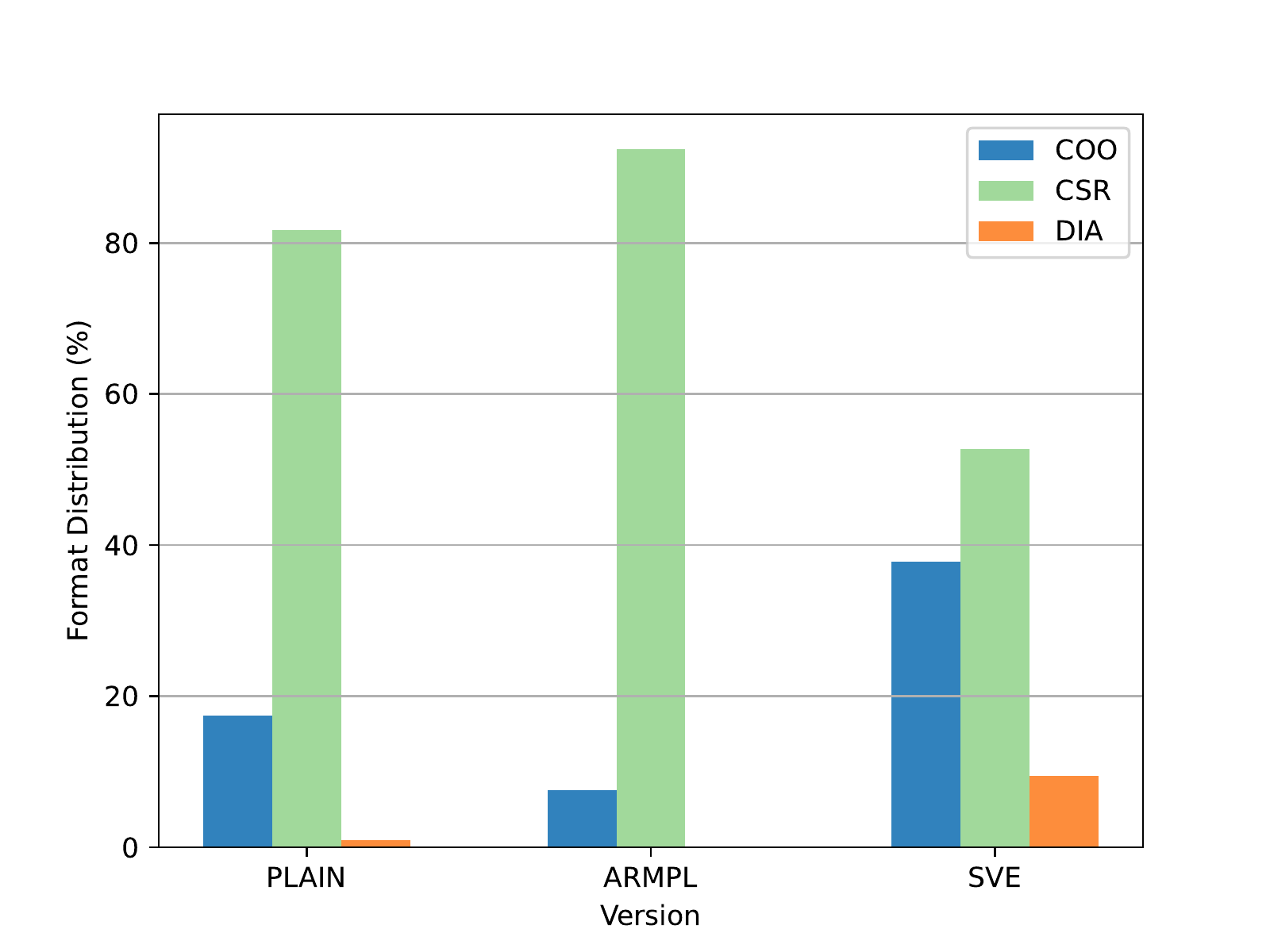}
    \setlength{\belowcaptionskip}{-8pt} 
    \caption{Distribution of the optimal format for the \gls{spmv} multiplication operation in serial for over 2100 sparse matrices from SuiteSparse collection on A64FX. Distributions are shown for each version of the algorithm.}
    \label{fig:distribution}
\end{figure}

The optimal format distribution per version differs significantly, as shown in Figure~\ref{fig:distribution}. For most of the matrices in the SuiteSparse collection the optimal format is \gls{csr}, validating its role as the most commonly used storage format. However, almost $20\%$ and $40\%$ of the matrices are better with \gls{coo} in the \emph{Plain} and \emph{SVE} versions respectively. Interestingly, although \gls{dia} format is almost of no use for \emph{Plain} version, the vectorization performed by \emph{SVE} version makes \gls{dia} format optimal for $10\%$ of the matrices. This indicates that the vectorization performed by the compiler for \gls{dia} in \emph{Plain} version might not be as effective as the use of custom \gls{sve} extensions in the \emph{SVE} version. The main takeaway here is that for the same hardware, operation and set of matrices in the majority of times the optimal performance is given by \gls{csr}, although the distribution of the optimal format can vary significantly given a different implementation or by applying different optimisations.

\begin{figure*}
    \centering
    \captionsetup{justification=centering}
    \begin{subfigure}[h]{0.32\textwidth}
            \captionsetup{justification=centering}
            \includegraphics[width=\columnwidth]{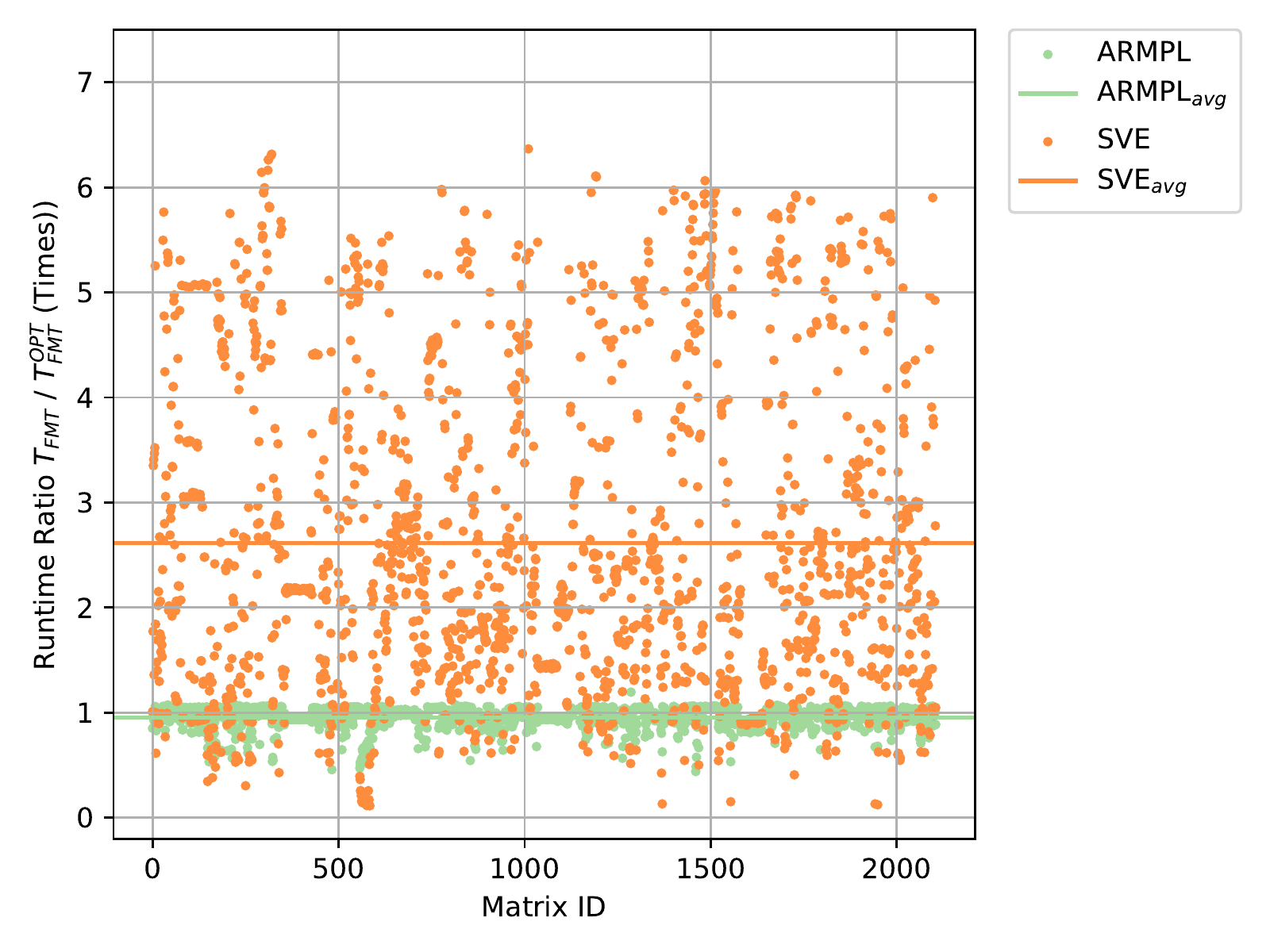}
            \caption{COO}
            \label{fig:arm-spmv-coo}
    \end{subfigure}
    ~
    \begin{subfigure}[h]{0.32\textwidth}
            \captionsetup{justification=centering}
            \includegraphics[width=\columnwidth]{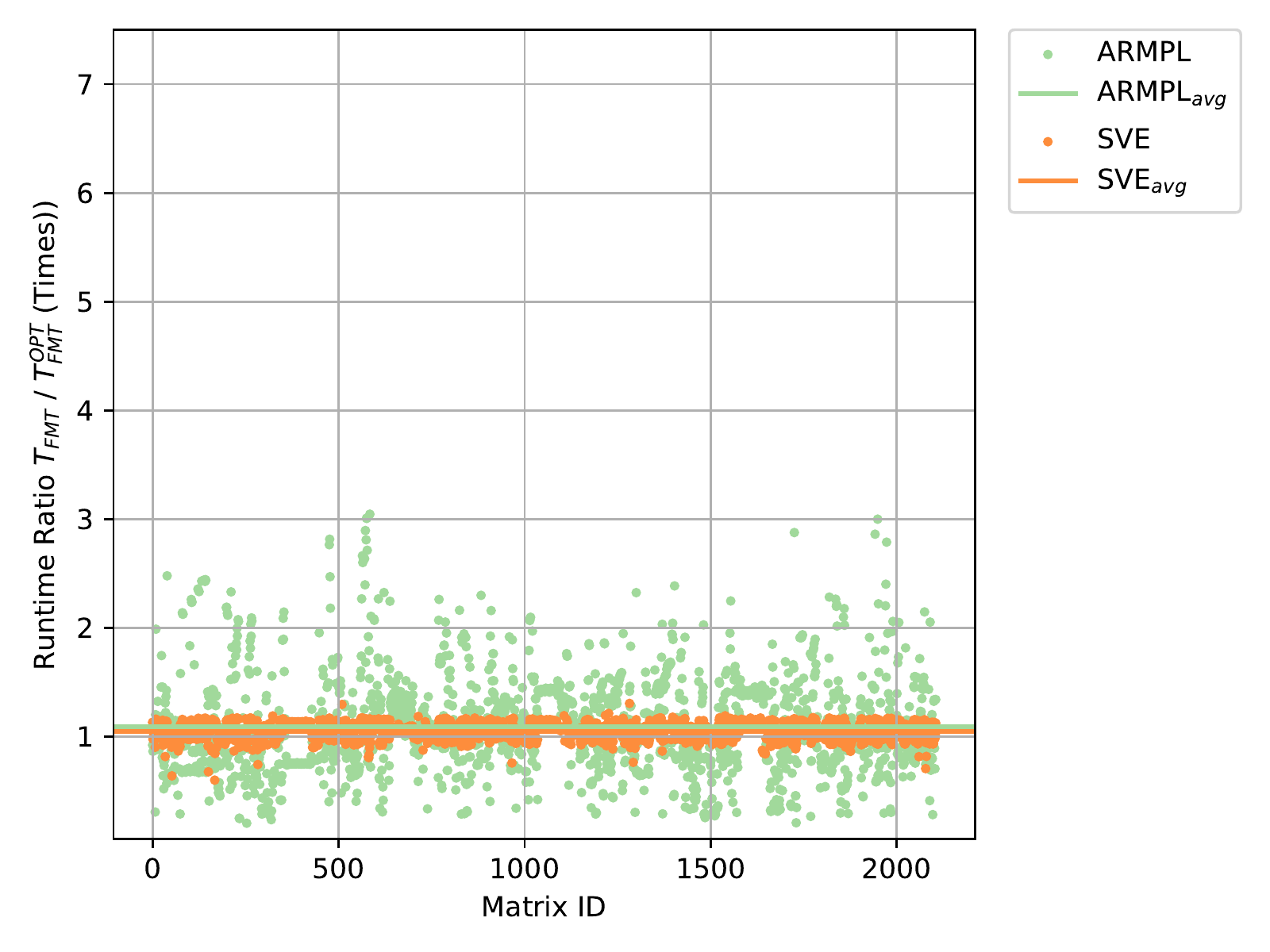}
            \caption{CSR}
            \label{fig:arm-spmv-csr}
    \end{subfigure}
    ~
    \begin{subfigure}[h]{0.32\textwidth}
            \captionsetup{justification=centering}
            \includegraphics[width=\columnwidth]{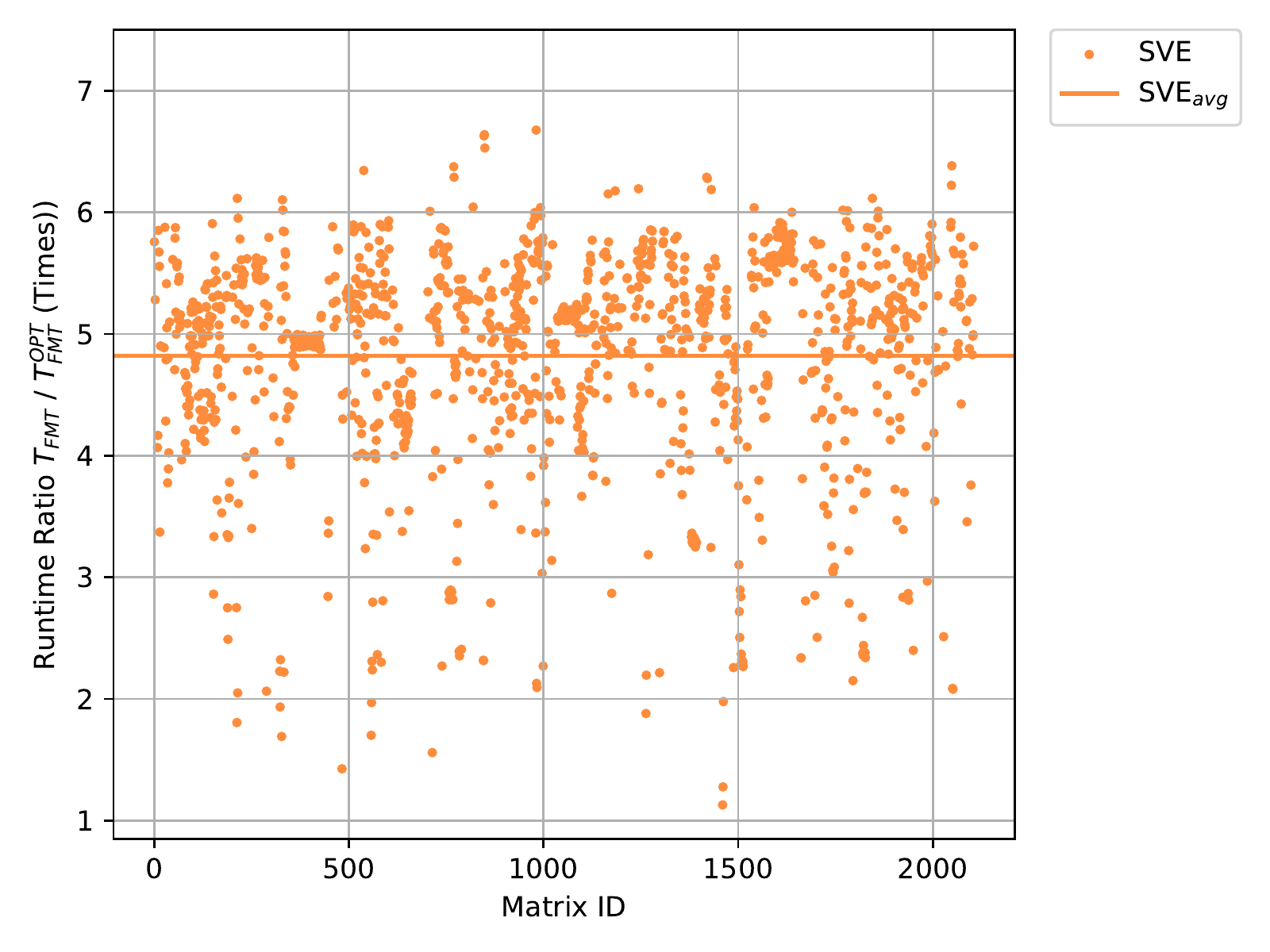}
            \caption{DIA}
            \label{fig:arm-spmv-dia}
    \end{subfigure}
    \caption{Serial performance of the \gls{spmv} multiplication over 2100 sparse matrices from SuiteSparse collection on A64FX. For each format, the original performance (\emph{Plain}) of the \emph{Morpheus} \gls{spmv} is measured against the optimized (ARM) \gls{spmv}. Optimized versions include the \emph{ArmPL} and \emph{SVE} implementations and the formats considered are COO, CSR, DIA. A ratio above $1$ indicates a speedup over the performance achieved when using the original implementation with the same format. The straight lines represent the average speedup over all matrices for each version.}
    \label{fig:arm-spmv}
    
    \begin{subfigure}[h]{0.42\textwidth}
            \captionsetup{justification=centering}
            \includegraphics[width=\columnwidth]{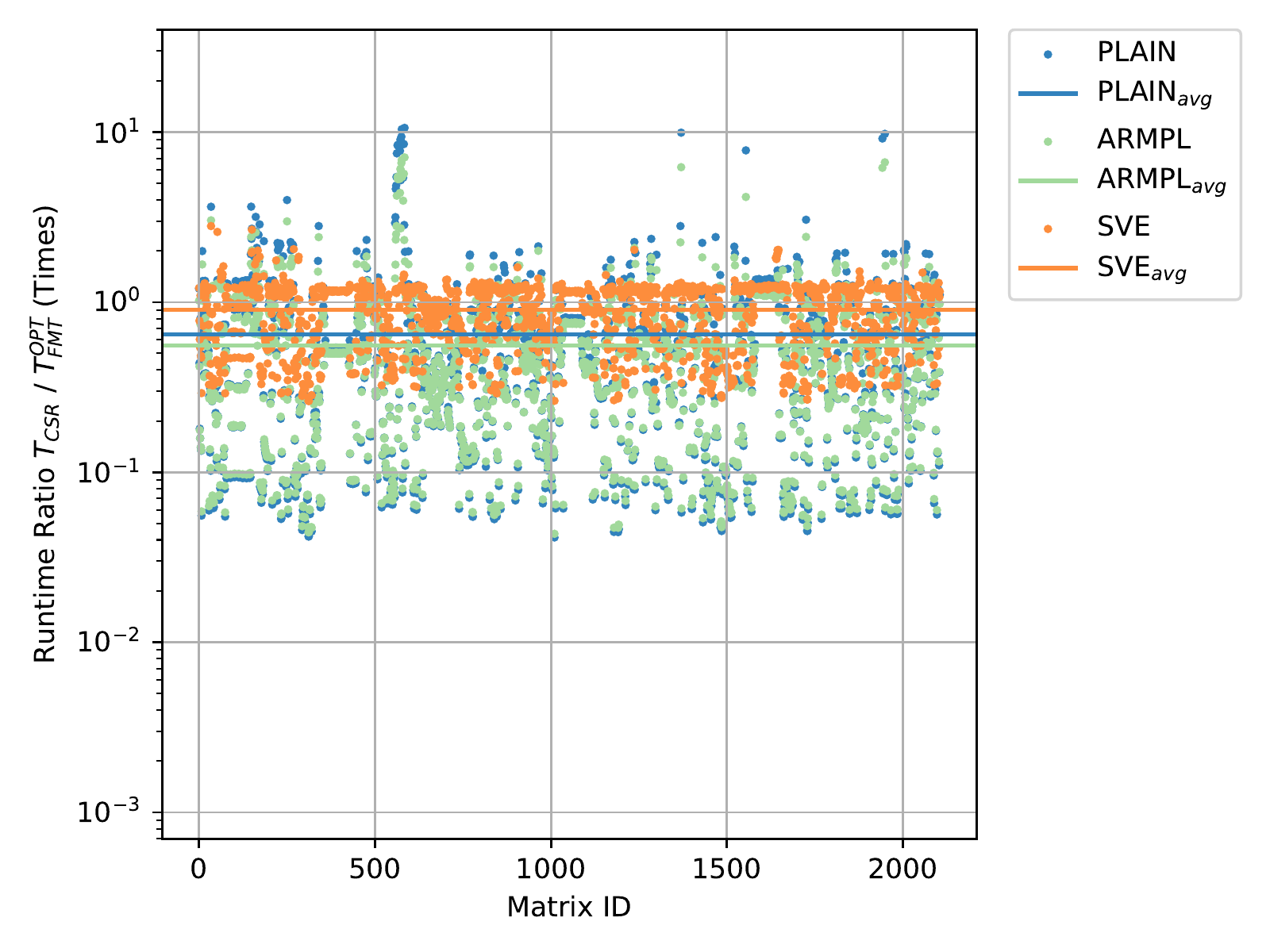}
            \caption{COO}
            \label{fig:arm-spmv-csr-coo}
    \end{subfigure}
    ~
    \begin{subfigure}[h]{0.42\textwidth}
            \captionsetup{justification=centering}
            \includegraphics[width=\columnwidth]{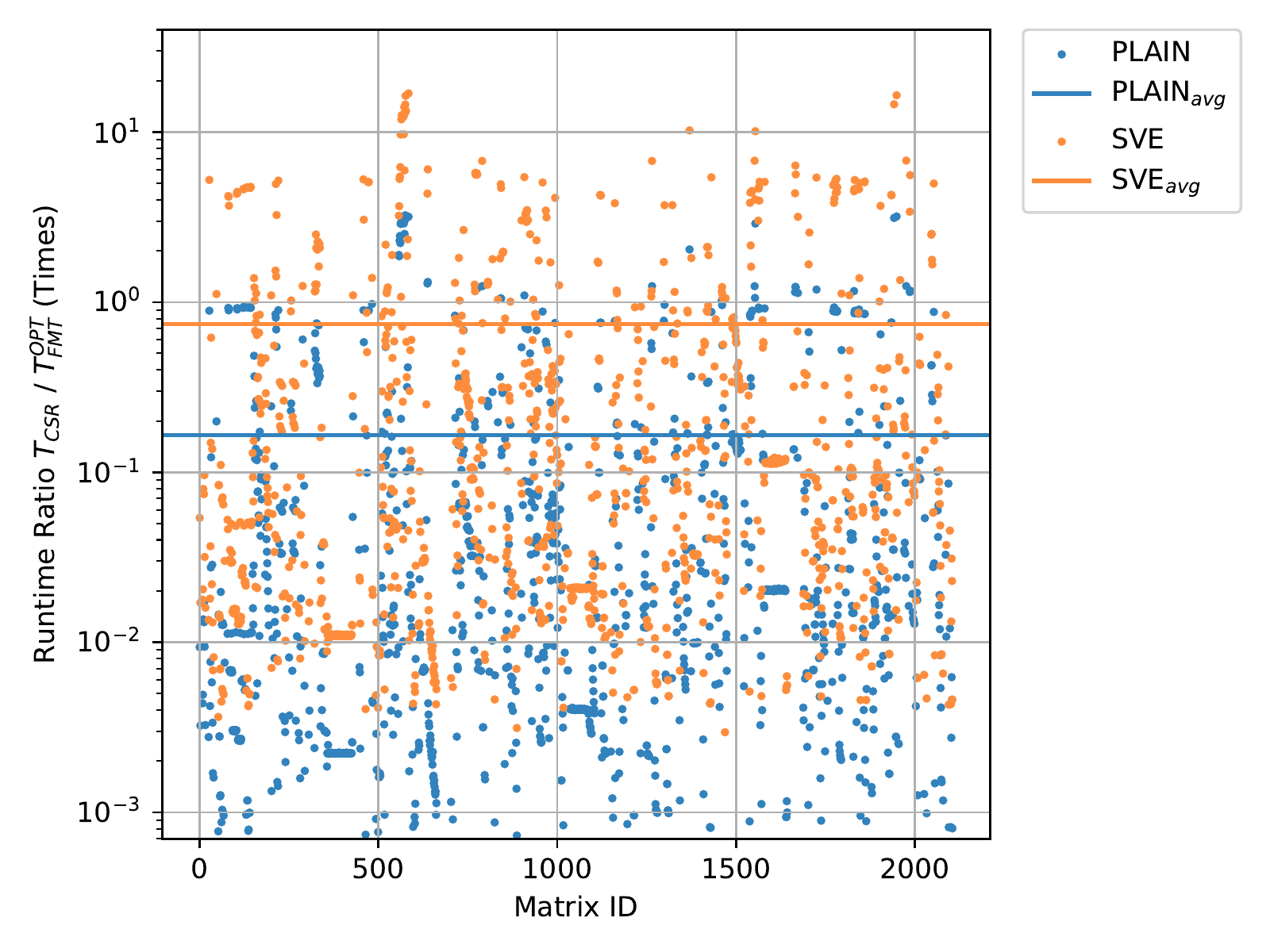}
            \caption{DIA}
            \label{fig:arm-spmv-csr-dia}
    \end{subfigure}
    \caption{Serial performance of the \gls{spmv} multiplication over 2100 sparse matrices from SuiteSparse collection on A64FX. For each format, the original performance (\emph{Plain}) of the \emph{Morpheus} CSR \gls{spmv} is measured against the optimized (ARM) \gls{spmv}. Optimized versions include the \emph{ArmPL} and \emph{SVE} implementations and the formats considered are COO, DIA. A ratio above $1$ indicates a speedup over the performance achieved when using the original CSR implementation. The straight lines represent the average speedup over all matrices for each version.}
    \label{fig:arm-csr-spmv}
\end{figure*}

Figure~\ref{fig:arm-spmv} shows the single-core performance of the \gls{spmv} multiplication for over 2100 sparse matrices from SuiteSparse collection on the A64FX processor. For each format, the runtime of the \emph{Plain} version \gls{spmv} is compared against the runtime of each optimized \gls{spmv} version (\emph{ARMPL} and \emph{SVE}) using the same format. For \gls{coo} (Figure~\ref{fig:arm-spmv-coo}), \emph{ARMPL} \gls{spmv} implementation performs at par with the \emph{Plain} \gls{coo} \gls{spmv} implementation whilst \emph{SVE} implementation consistently outperforms it, obtaining average speedups of $1\times$ and $3.6\times$ respectively. The increase in performance achieved by the \emph{SVE} version can be attributed in assumptions made during the implementation of the \gls{spmv} algorithm that allowed us to take advantage of different intrinsic commands. For example, by assuming that the matrix is sorted (which \emph{Morpheus} ensures prior to applying any \gls{spmv} operation) a tree-based reduction was used instead of the traditional left-to-right reduction in order to accumulate the results in the output vector~\texttt{y}. It is worth highlighting that even-though the \emph{SVE} version significantly outperforms the \emph{Plain} version for most of the matrices in \gls{coo}, there is still a noticeable number of matrices for which it significantly under-performs. For very sparse and unstructured matrices, \emph{SVE} version seems to introduce more overheads from the vectorization process effectively hindering the performance of \gls{spmv}. For \gls{csr} (Figure~\ref{fig:arm-spmv-csr}), the average runtime performance for both \emph{ARMPL} and \emph{SVE} versions is at par with \emph{Plain}. Interestingly, for a large number of matrices the \emph{ARMPL} version achieves speedups above $1\times$ compared to \emph{Plain}, with max speedup up to $3\times$. At the same time, for a large number of matrices it seems to significantly under-perform, whilst \emph{SVE} version offers a more stable performance profile. The largest benefit from exploiting the \gls{sve} intrinsics is reaped by \gls{dia} (Figure~\ref{fig:arm-spmv-dia}) where the \emph{SVE} version obtains an average speedup of $\approx5\times$ compared to the \emph{Plain} implementation. The fact that the \emph{SVE} implementation beats the \emph{Plain} implementation for all matrices in the set suggests that the compiler has a tough time performing effecting vectorization in \emph{Plain} version.

Since for most of the matrices the optimal format for performing the \gls{spmv} operation is \gls{csr}, in Figure~\ref{fig:arm-csr-spmv} we compare the runtime performance of \gls{coo} and \gls{dia} for all three versions (\emph{Plain}, \emph{ARMPL} and \emph{SVE}) against the runtime of the \gls{csr} for the \emph{Plain} version, with the same configuration as before. For \gls{coo} (Figure~\ref{fig:arm-spmv-csr-coo}), on average both \emph{Plain} and \emph{ARMPL} versions perform approximately the same and do worse compared to the \emph{Plain} \gls{csr} implementation. On the other hand, the optimisations performed in the \emph{SVE} version achieve an average performance at par compared to the \emph{Plain} \gls{csr} implementation. Note that for most of the matrices, \gls{coo} will result in significant slowdowns compared to \gls{csr} irrespective of the version used. However, for a small number of matrices, all three versions offer noticeable speedups reaching up to max speedups of $10\times$. Similar observations can be made about \gls{dia} (Figure~\ref{fig:arm-spmv-csr-dia}). It is obvious that \gls{dia} format finds use in a small number of matrices with specific characteristics. However, for those matrices the performance optimisations from the adoption of \gls{sve} intrinsics can offer a significant boost in performance with max speedups of $\approx20\times$ compared to \gls{csr}.

It is worth pointing out, that even-though it was expected for the \emph{ARMPL} versions to perform optimally for \gls{coo} and \gls{csr}, on average they were at par with their \emph{Plain} equivalents. However, for a large number of matrices, the \gls{csr} \emph{ARMPL} implementation was optimal. Furthermore, the adoption of \gls{sve} intrinsics had a noticeable impact on \gls{dia}, increasing the number of matrices it was optimal for by an order of magnitude and offering consistently noticeable speedups compared to it's equivalent \emph{Plain} implementation. This experiment makes it clear that in the same way no single format can perform best across the different sparsity patterns, no single implementation can do the same either. As a result, these findings motivate the extension of \emph{Morpheus} to also support an efficient mechanism for selecting the optimal implementation at runtime. 

% \textbf{Remarks:}
% \begin{itemize}
%     \item Fig.\ref{fig:arm-spmv-coo}: ARMPL COO significantly underperforms compared to original (plain) COO but SVE much better.
%     \begin{itemize}
%         \item Probably implementation doesn't assume matrix is sorted.
%         \item Our SVE implementation assumes it - average speedup $2\times$.
%         \item However, for a set of matrices that are generally very sparse, SVE adds more overheads resulting in lower max speedups compared to plain COO (see Fig~\ref{fig:arm-spmv-csr-coo})
%         \item COO SVE consistently outperforms CSR plain (Fig~\ref{fig:arm-spmv-csr-coo}) - not the case before.
%     \end{itemize}
%     \item Optimized CSR versions consistently outperform the plain CSR but for some matrices ARMPL performance drastically drops.
%     \item Overall DIA underperforms (plain) or at par (SVE) wrt to CSR (Fig.~\ref{fig:arm-spmv-csr-dia}
%     \begin{itemize}
%         \item However, for $\approx10\%$ of the matrices, DIA significantly outperforms CSR (plain) with max speedups reaching up to $30\times$ for SVE.
%         \item DIA is significantly benefited from vectorization - suitable candidate for SVE optimisation.
%         \item Optimized mplementations though come with limitations i.e currently only work for layout right matrices.
%     \end{itemize}
% \end{itemize}

\begin{figure*}
    \centering
    %\captionsetup{justification=centering}
    %\begin{subfigure}[h]{0.32\textwidth}
    %        \captionsetup{justification=centering}
    %        \includegraphics[width=\columnwidth]{figures/spmv/opt-fpga/stemplot_CSR_speedup_COO.pdf}
    %        \caption{COO}
    %        \label{fig:fpga-spmv-coo}
    %\end{subfigure}
    %~
    %\begin{subfigure}[h]{0.32\textwidth}
    %        \captionsetup{justification=centering}
    %        \includegraphics[width=\columnwidth]{figures/spmv/opt-fpga/stemplot_speedup_CSR.pdf}
    %        \caption{CSR}
    %        \label{fig:fpga-spmv-csr}
    %\end{subfigure}
    %~
    %\begin{subfigure}[h]{0.32\textwidth}
    %        \captionsetup{justification=centering}
    %        \includegraphics[width=\columnwidth]{figures/spmv/fpga/stemplot_speedup_DIA.pdf}
    %        \caption{DIA}
    %        \label{fig:fpga-spmv-dia}
    %\end{subfigure}
    %\caption{Serial performance of the \gls{spmv} multiplication over 2100 sparse matrices from SuiteSparse collection on A64FX. For each format, the original performance (\emph{Plain}) of the \emph{Morpheus} \gls{spmv} is measured against the optimized (ARM) \gls{spmv}. Optimized versions include the \emph{ArmPL} and \emph{SVE} implementations and the formats considered are COO, CSR, DIA. A ratio above $1$ indicates a speedup over the performance achieved when using the original implementation with the same format. The straight lines represent the average speedup over all matrices for each version.}
    %\label{fig:fpga-spmv}

    \begin{subfigure}[h]{0.32\textwidth}
            \captionsetup{justification=centering}
            \includegraphics[width=\columnwidth]{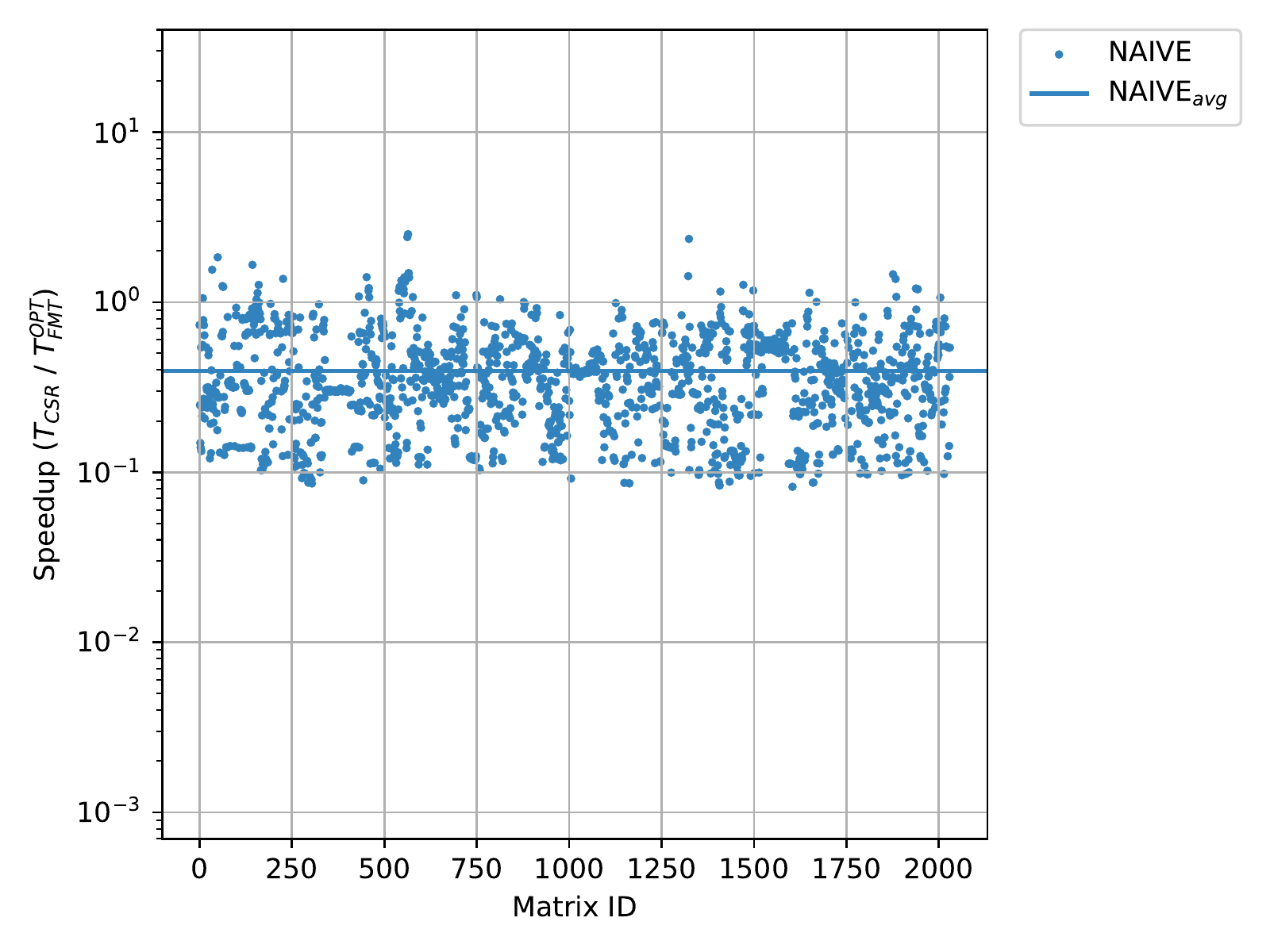}
            \caption{Plain \gls{coo} \gls{spmv} performance against \gls{csr} implementation.}
            \label{fig:fpga-spmv-csr-coo}
    \end{subfigure}
    ~
    \begin{subfigure}[h]{0.32\textwidth}
            \captionsetup{justification=centering}
            \includegraphics[width=\columnwidth]{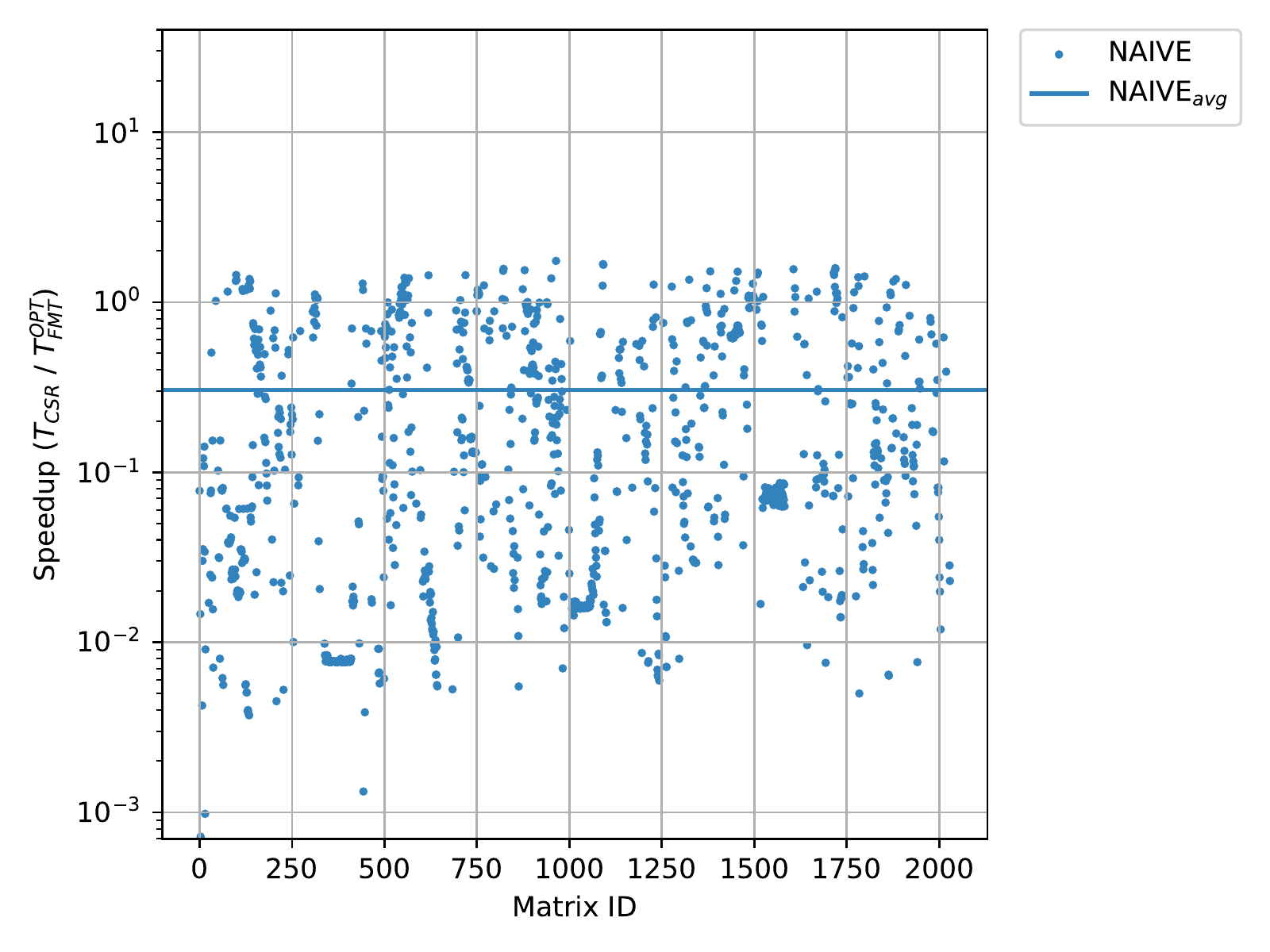}
            \caption{Plain \gls{dia} \gls{spmv} performance against \gls{csr} implementation}
            \label{fig:fpga-spmv-csr-dia}
    \end{subfigure}
    ~
    \begin{subfigure}[h]{0.32\textwidth}
            \captionsetup{justification=centering}
            \includegraphics[width=\columnwidth]{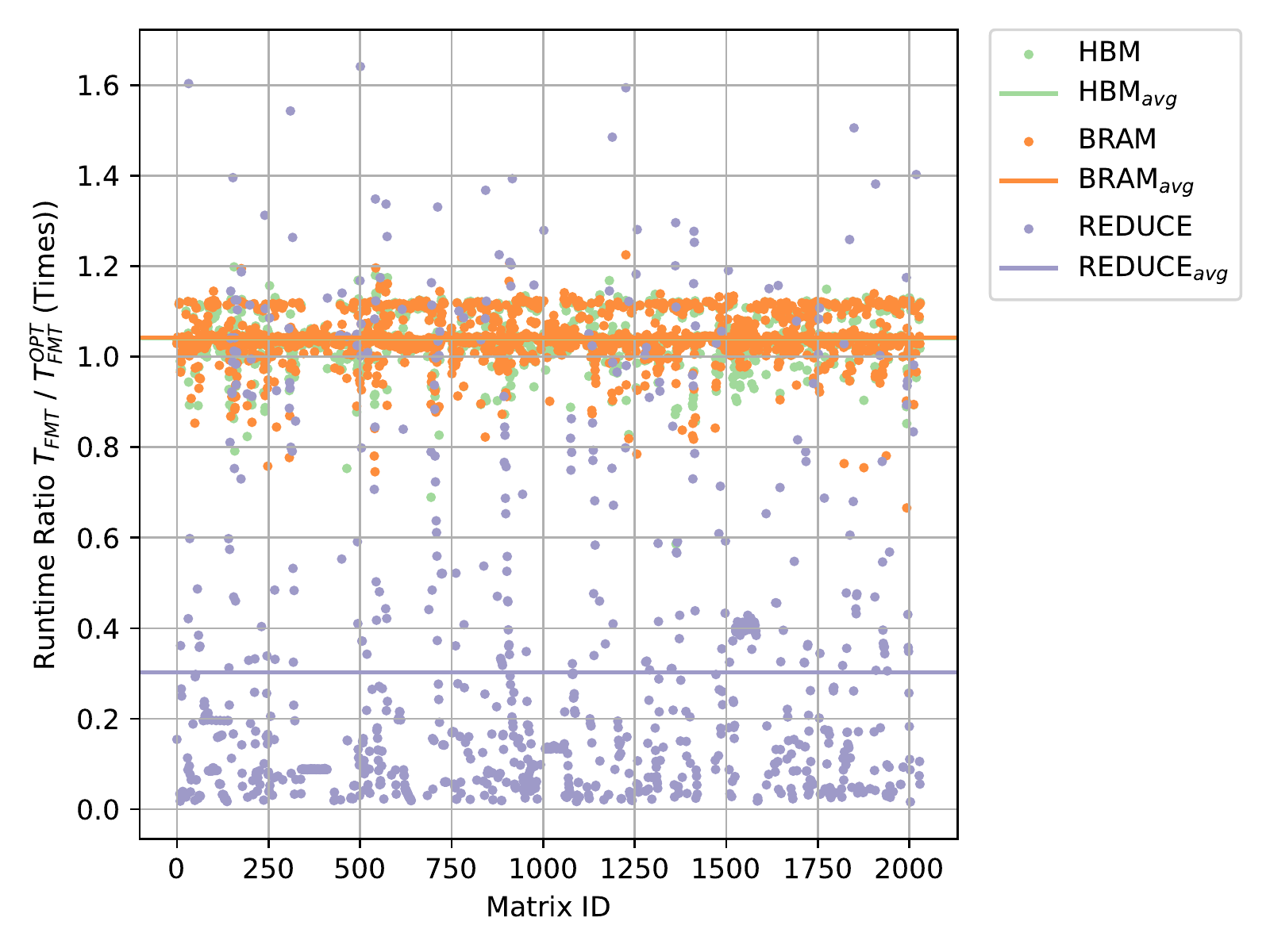}
            \caption{Optimised versions of \gls{coo} \gls{spmv} against plain \gls{coo} version.}
            \label{fig:fpga-spmv-coo}
    \end{subfigure}
    \caption{Serial performance of the \gls{spmv} multiplication over 2100 sparse matrices from SuiteSparse collection on Alveo U280. %For each format, the original performance (\emph{Plain}) of the \emph{Morpheus} CSR \gls{spmv} is measured against the optimized (ARM) \gls{spmv}. Optimized versions include the \emph{ArmPL} and \emph{SVE} implementations and the formats considered are COO, DIA. 
    A ratio above $1$ indicates a speedup over the performance achieved against a reference implementation. The straight lines represent the average speedup over all matrices for each version.}
    \label{fig:fpga-csr-spmv}
\end{figure*}
\subsection{Evaluation of \gls{spmv} on FPGAs}

\begin{figure}[h]
    \centering
    \includegraphics[width=0.8\columnwidth]{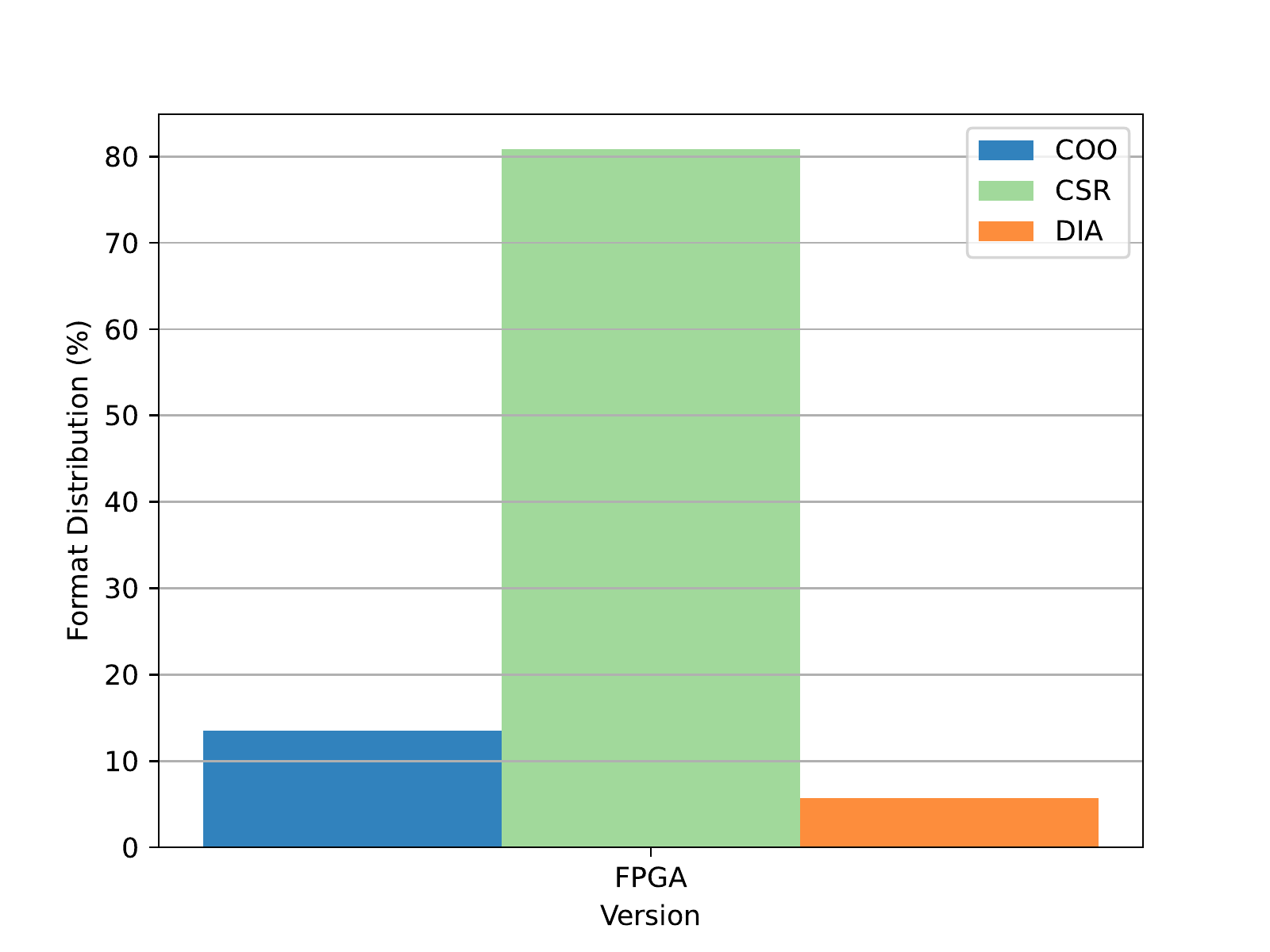}
    \setlength{\belowcaptionskip}{-8pt} 
    \caption{Distribution of the optimal format for the \gls{spmv} multiplication operation in serial for over 2100 sparse matrices from SuiteSparse collection on Alveo U280. Distributions are shown for each version of the algorithm.}
    \label{fig:distribution_fpga}
\end{figure}

When evaluating our FPGA implementations of \gls{spmv}, we performed 10 iterations of each of the three \gls{spmv} kernels over %\textcolor{red}{xxx} sparse matrices from % \ref{sec:fpga_limiations} 
the SuiteSparse collection, with Section~\ref{sec:setup} providing more details on the experimental setup. Figure~\ref{fig:distribution_fpga} shows the distribution of optimal formats for the \gls{spmv} operation in Serial for over 2100 sparse matrices. The \gls{csr} format dominates in terms of performance or shortest runtime, since for more than 80\% of the matrices performs optimally when compared against the \gls{coo} and \gls{dia} format. The second most optimal format across the SuiteSparse matrices is COO with more than 10\% and DIA is optimal still for more than 5\% of the matrices. It is worth highlighting that on the A64FX processor, similar distribution was obtained for the \emph{Plain} \gls{spmv} Version as shown in Figure~\ref{fig:distribution}.%compare against AArch64

The serial performance of our FPGA prototypes for COO and DIA against \gls{csr} is presented in Figures~\ref{fig:fpga-spmv-csr-coo} and \ref{fig:fpga-spmv-csr-dia}, confirming the conclusions from the optimal format comparison where the SpMV algorithm for CSR outperforms both the COO and DIA implementations on FPGA for the majority of matrices. Note that even-though the average speedup for both \gls{coo} and \gls{dia} is well below 1 when compared against \gls{csr}, for a very few matrices we do observe some increase in performance (for instance, in one case with \gls{coo} we observe a $2\times$ speedup). Optimising the kernels further such that they are exploiting optimally the characteristics of the hardware can have the potential of a more diverse distribution of optimal formats. However, at this point with \gls{coo} and \gls{dia} both yielding no speedup over \gls{csr}, we can report that the compressed sparse row advantage on traditional architectures also holds true for our baseline versions on FPGA. %compare against AArch64

In Figure~\ref{fig:fpga-spmv-coo} the optimisations of the \gls{coo} \gls{spmv} kernel described in Section~\ref{sec:spmvFPGA} are compared against the \emph{naive} version. We observed that the use of \gls{hbm} provided a small boost in performance for most of the matrices, but combining HBM with on-chip BRAM had no further noticeable effect. The \emph{reduce} operation, shown in Figure~\ref{fig:fpga_coo_dataflow_region}, is sub-optimal for FPGAs because the floating point accumulation between iterations, which requires more than one cycle, adds a spatial dependency between loop iterations and thus pushes the initiation interval higher than one, meaning that cycles are wasted. It is common practice on FPGAs to optimise this by reducing in chunks of independent iterations, and performance numbers for this are reported in Figure~\ref{fig:fpga-spmv-coo} by \emph{REDUCE}. However it can be seen that, in this case, this is not beneficial and that is because the conditional on \textit{row index} means that the cycles which do not match the conditional are \textit{doing nothing} anyway, and-so this offsets the added complexity of the reduction optimisation. However, it is worth highlighting that it achieved the highest maximum speedups out of all three optimisations for one of the matricies, where we observe that the reduce optimisation works better for small matrices as the overhead increases with larger \texttt{nrows} and \texttt{NNZ}. This result demonstrates once more, this time on FPGAs, the importance of having multiple implementations of the same \gls{spmv} operation and the ability to dynamically switch to the optimal format given the sparsity pattern of the input.
% \ref{sec:fpga_limiations} % single buffer size limit (4GB) and accumulated limit across input buffers (8GB HBM)

% optimal format distribution: CSR dominant (same as for CPU/AArch64
% check and compare dataflowed/optimised perf

% explain Figures for serial performance

\begin{figure*}
    \centering
    \begin{subfigure}[h]{0.32\textwidth}
            \captionsetup{justification=centering}
            \includegraphics[width=\columnwidth]{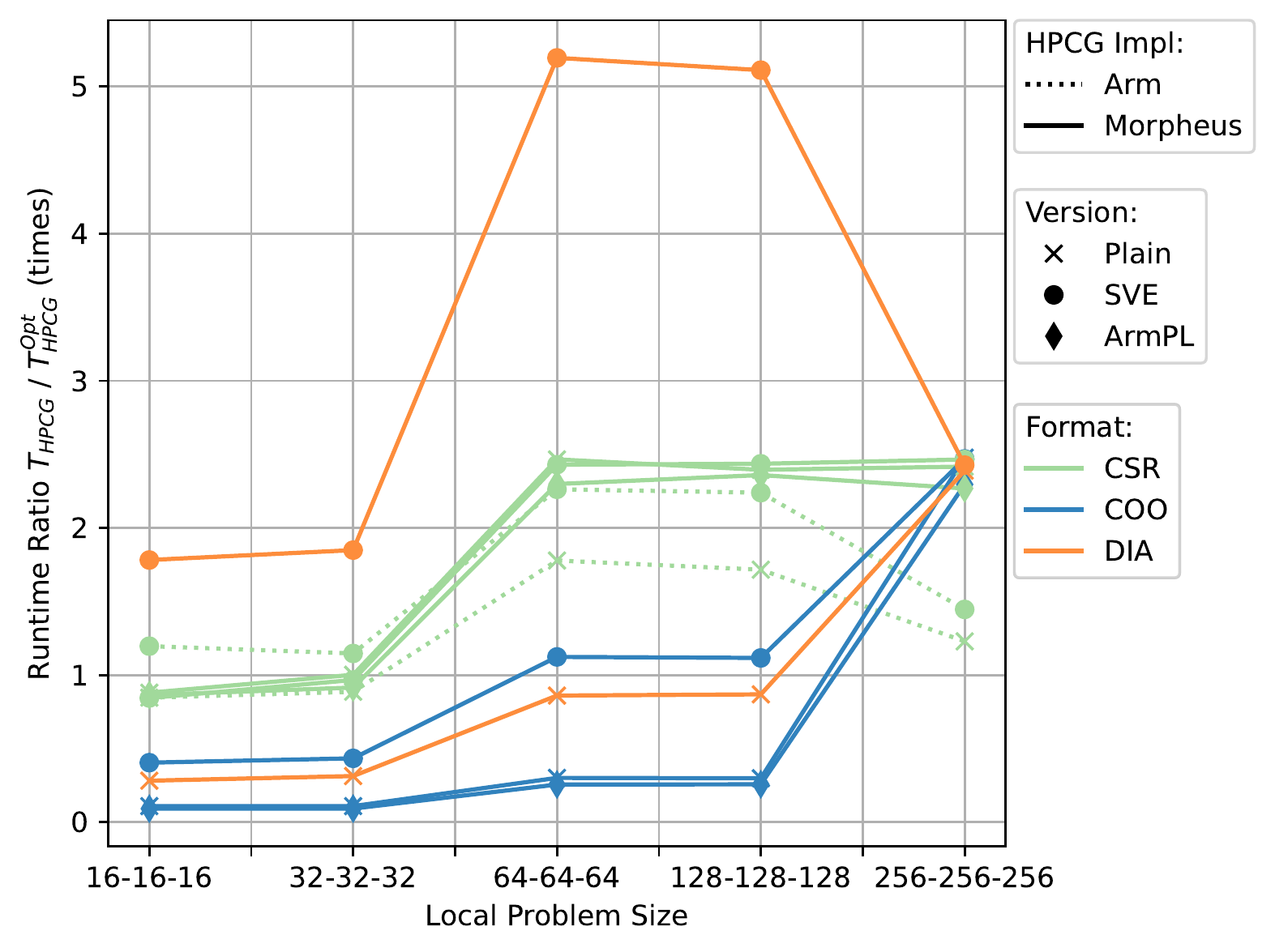}
            \caption{Serial Performance}
            \label{fig:serial-hpcg}
    \end{subfigure}
    ~
    \begin{subfigure}[h]{0.32\textwidth}
            \captionsetup{justification=centering}
            \includegraphics[width=\columnwidth]{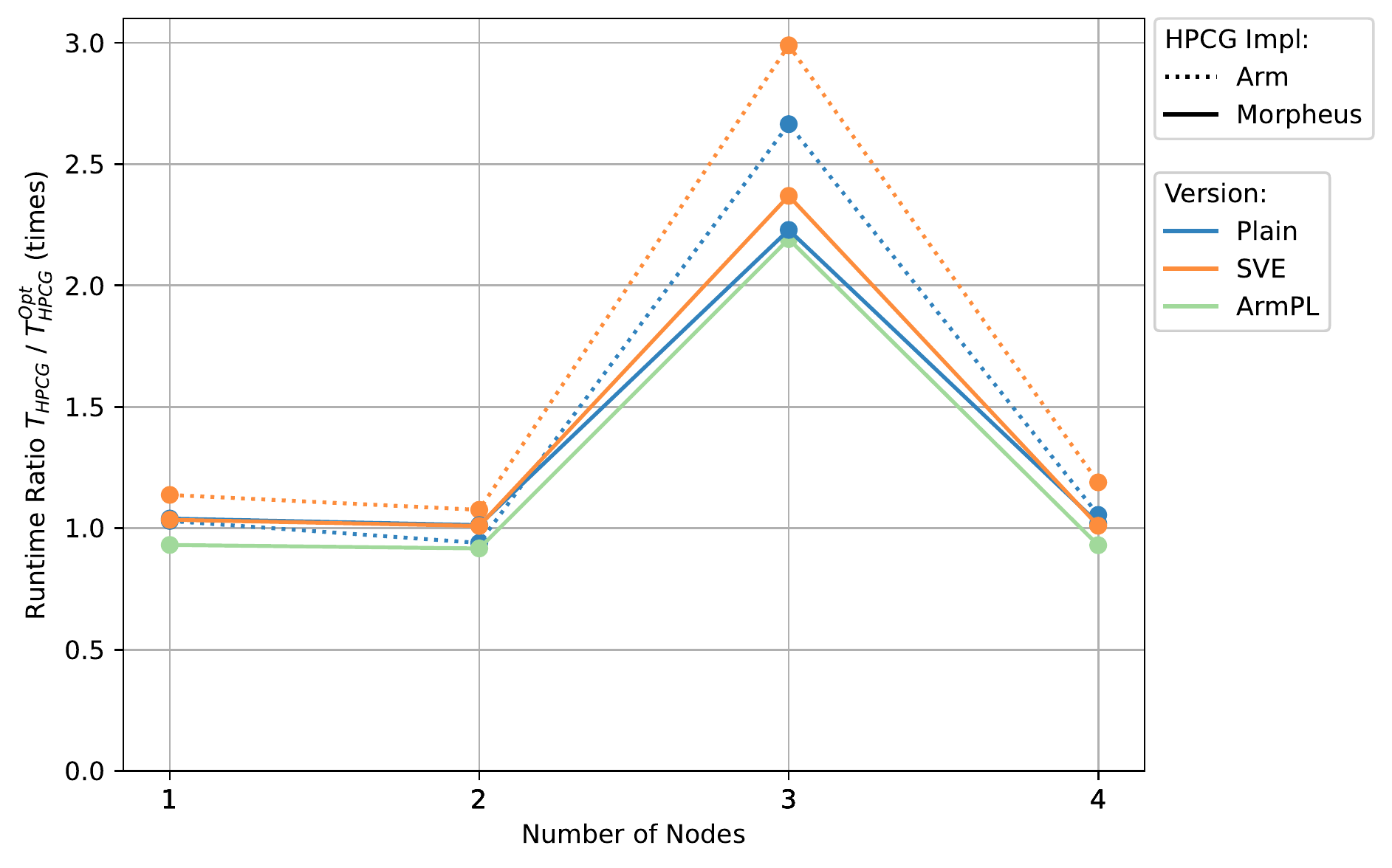}
            \caption{Strong Scaling}
            \label{fig:arm-hpcg-strong}
    \end{subfigure}
    ~
    \begin{subfigure}[h]{0.32\textwidth}
            \captionsetup{justification=centering}
            \includegraphics[width=\columnwidth]{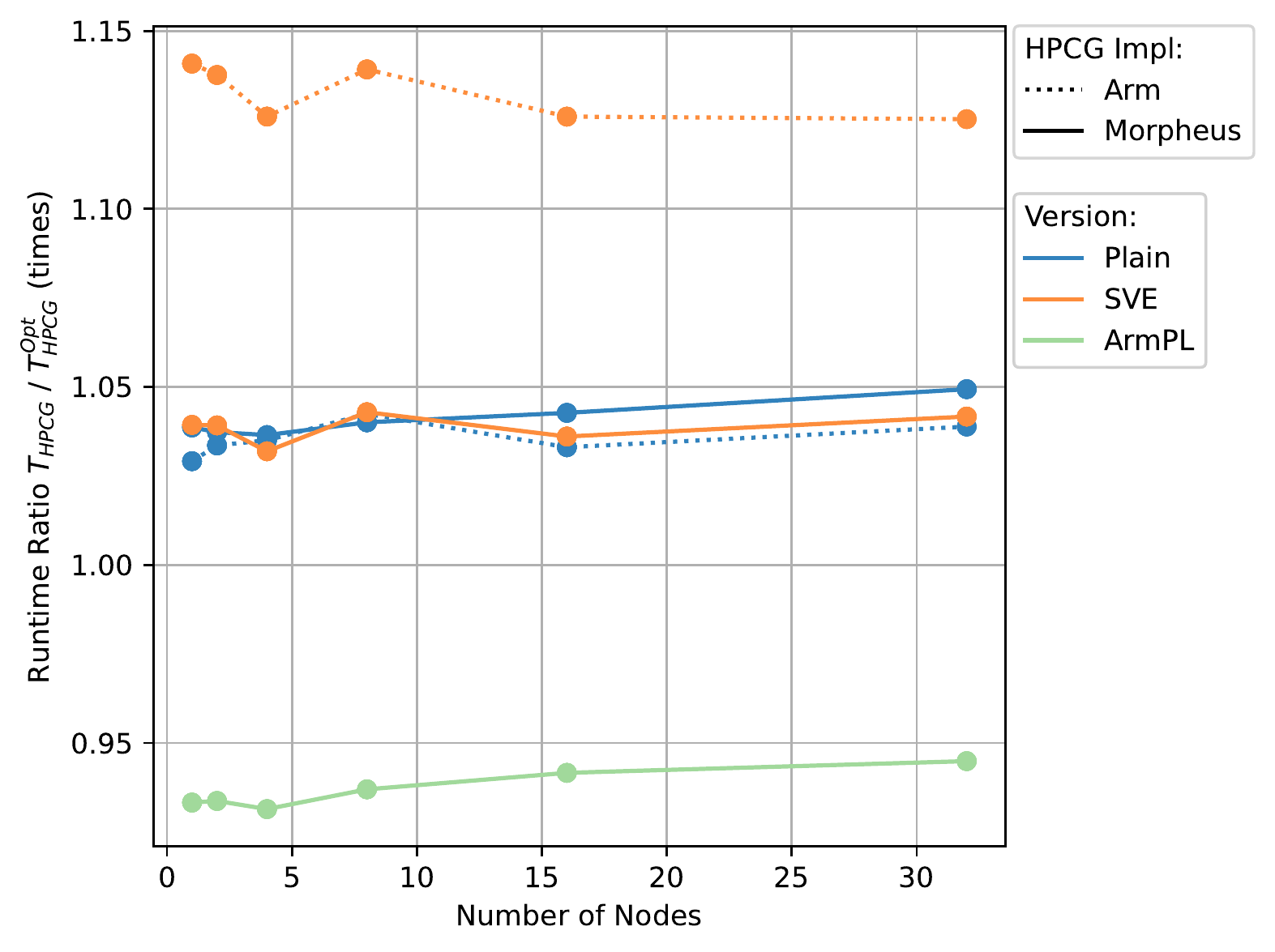}
            \caption{Weak Scaling}
            \label{fig:arm-hpcg-weak}
    \end{subfigure}
    \caption{Performance of the Morpheus- and Arm-enabled HPCG implementations. The performance is measured as the \gls{spmv} runtime ratio of the reference HPCG w.r.t each optimal HPCG implementation and version on A64FX. For the serial version, \gls{hpcg-bench} runs for a set of problem sizes. For the distributed Morpheus-enabled HPCG a run-first auto-tuner is used in order to determine the optimal format to use at each process. A ratio above 1 indicates a speedup over the performance achieved when using the original HPCG.}
    \label{fig:arm-scaling}
\end{figure*}

\subsection{HPCG experiments}
The \gls{hpcg-bench} benchmark solves the Poisson differential equation on a regular 3D grid, discretized with a 27-point stencil. It uses the \gls{pcg} algorithm with a symmetric Gauss-Seidel\cite{Saad_2003} as a preconditioner, and includes the following computations: sparse matrix-vector multiplications (\glspl{spmv}); vector updates; global dot products; a local symmetric Gauss-Seidel smoother (including a sparse triangular solve); and multi-grid (MG) preconditioned solvers. The performance bottleneck in \gls{hpcg-bench} is due to to the sparse operations that are carried out at every step of the iterative solver, i.e the \gls{spmv} and the Gauss-Seidel smoother.

\gls{hpcg-bench} is a widely accepted and well-understood benchmark used to measure the performance of \gls{hpc} systems. For this reason, multiple vendor specific implementations exist (\cite{cuda-hpcg,arm-hpcg,xilinx-hpcg}). In previous work, we have also implemented a \emph{Morpheus}-enabled \gls{hpcg-bench}\cite{morpheus-hpcg}. The benchmark progresses in the following phases:
\begin{enumerate}
    \item \textit{Problem setup}: Constructs the synthetic problem by creating the geometry and linear system.
    \item \textit{Reference timing}: Measures the time taken to run the \gls{spmv} and MG reference implementations and the time to solution for the reference \gls{cg} solver.
    \item \textit{Problem Optimisation setup}: Configures the user defined data structures to be used in the optimised problem.
    \item \textit{Validation and Verification}: Checks that the optimised problem has returned the expected results.
    \item \textit{Optimised problem timing}: Measures the time to solution for the optimised \gls{cg} solver.
\end{enumerate}

In the following experiment, we benchmark the performance of: 1)~the \emph{Morpheus}-enabled \gls{hpcg-bench} with the newly added \emph{ARMPL} and \emph{SVE} versions of \gls{spmv} and 2)~The vendor (Arm) implementation of \gls{hpcg-bench} against the original \gls{hpcg-bench}. We are focusing on Phase~5, although for the purposes of this work, since we are interested in the \gls{spmv} multiplication, we are disabling the use of the preconditioner from all implementations. The experiment is configured as described in Section~\ref{sec:setup} on the A64FX processors.

%\subsection{Serial HPCG}
In Figure~\ref{fig:serial-hpcg}, the single-core \gls{spmv} runtime performance of the two \gls{hpcg-bench} implementations (\emph{Morpheus} and \emph{Arm}) is measured against the original \gls{hpcg-bench} over a set of different problem sizes. For each format in the \emph{Morpheus}-enabled implementation, we measure the runtime for the \emph{Plain}, \emph{ARMPL} and \emph{SVE} versions of the \gls{spmv} multiplication routines. In a similar way, the runtime of the \gls{spmv} with (\emph{SVE}) and without (\emph{Plain}) \gls{sve} intrinsics is measured for the Arm implementation.

% \begin{figure}[t]
%     \centering
%     \includegraphics[width=0.9\columnwidth]{figures/serial-hpcg/hpcg-serial-TimeSpmv_speedup.pdf}
%     \setlength{\belowcaptionskip}{-8pt} 
%     \caption{Serial performance of the Morpheus- and Arm-enabled HPCG implementations. The performance is measured as the SpMV runtime ratio of the reference HPCG w.r.t each optimal HPCG implementation, version and format over a set of problem sizes on A64FX. A ratio above 1 indicates a speedup over the performance achieved when using the original HPCG.}
%     \label{fig:serial-hpcg}
% \end{figure}

The system matrix in \gls{hpcg-bench} is generated using the \gls{fdm}. As a result, the matrix is highly regular with non-zeros around the diagonals. It is expected therefore that the \gls{dia} format would perform optimally compared to the rest of the formats, a hypothesis which is confirmed by Figure~\ref{fig:serial-hpcg}. The optimal performance difference between the two versions of the \gls{dia} \gls{spmv} i.e \emph{Plain} and \emph{SVE}, follows the average performance observed in Figure~\ref{fig:arm-spmv-csr-dia}, with a max speedup of $5\times$ compared to the reference \gls{hpcg-bench}. Note that for smaller problem sizes, the performance of both is impacted by the extra operations due to zero-padding. The \emph{SVE} version for the Arm \gls{hpcg-bench} closely follows the performance of the \emph{Morpheus}-enabled \gls{hpcg-bench} that uses \gls{csr}. However, the performance of the Arm \gls{hpcg-bench} version without \emph{SVE} support diminishes at the problem size of $64^3$. This can be attributed to the fact the matrix in Arm \gls{hpcg-bench} is reordered and to the lack of \gls{sve} extensions. Interestingly, for a problem size of $256^3$ every implementation that was performing better, compared to the reference, now either sees a drop in performance or stays at par. However, implementations such as all versions of the \emph{Morpheus}-enabled \gls{hpcg-bench} using \gls{coo} and the \emph{Plain} version that uses \gls{dia} now see a boost in performance, with the \emph{SVE} version of \gls{coo} achieving the optimal --but marginal-- performance out of all. This result further motivates the need for runtime switching of different \gls{spmv} implementations for the same format.

%\subsection{Multi-node HPCG}
The performance of the distributed \gls{hpcg-bench} for both Morpheus- and Arm-enabled \gls{hpcg-bench} is measured against the original \gls{hpcg-bench} implementation. For the strong scaling experiment the global problem size chosen is $192\times256\times192$ and for the weak scaling experiment the local problem size is $48\times64\times64$. For the distributed implementations the sparsity pattern of the matrix on each process differs from the one in the Serial case, due to the remote elements of the matrix added to the right. Whilst the matrix is initially structured, the remote part of it is highly unstructured. As a result, in the \emph{Morpheus}-enabled \gls{hpcg-bench} we physically split this matrix into \emph{local} and \emph{remote} part in order to potentially select different storage formats for each. This is achieved by utilising a run-first auto-tuner where it finds the optimal format to use on every process. The formats selected on each process for each version of our implementation are shown in Table~\ref{tab:scaling-formats}. Notice that for the \emph{SVE} version, the optimal formats chosen were \gls{dia} and \gls{coo}, for both strong and weak scaling experiments.

\begin{table}[h]
    \centering
    \caption{Optimal format used for each version of the \emph{Morpheus}-enabled HPCG across the different processes for both strong and weak scaling.}
    \label{tab:scaling-formats}
    \begin{tabular}{c|c|c|c}
        Version & Local Format & Remote Format \\
        \hline
        Plain & \gls{csr} & \gls{csr} \\
        ARMPL & \gls{csr} & \gls{csr} \\
        SVE   & \gls{dia} & \gls{coo} \\
    \end{tabular}
\end{table}

Figures~\ref{fig:arm-hpcg-strong} and \ref{fig:arm-hpcg-weak} shows the scaling \gls{spmv} performance for each version of \emph{Morpheus}- and Arm-enabled \gls{hpcg-bench} implementations against the reference \gls{hpcg-bench}. Note that for both strong and weak scaling, the \emph{Morpheus}-enabled \gls{hpcg-bench} closely tracks the performance of the optimal \emph{SVE} version of Arm-enabled \gls{hpcg-bench}. In Figure~\ref{fig:arm-hpcg-strong}, the boost in performance on 3~nodes happens due to the fact the size of the system matrix strikes a balance between the benefits achieved from vectorisation and the overheads that are associated with it. Furthermore, the weak scaling results in Figure~\ref{fig:arm-hpcg-weak}, even-though they show marginal improvement over the original \gls{hpcg-bench}, still show that our contributions match the performance of the optimal Arm implementation. Note however, with a local problem size somewhere closer to the region we have previously noticed performance improvements (i.e between $64^3$ and $128^3$ as shown in Figure~\ref{fig:serial-hpcg}) more noticeable speedups would have been expected, although due to memory limitations such runs weren't feasible at the distributed level.

%% file: src/related_work.tex
\section{Related Work}

The research efforts into optimising sparse computations largely fall in two categories, namely (i) in the creation of novel storage formats that better capture the characteristics of particular architectures or sparsity patterns, such as CSR5~\cite{csr5} and Sell-C-$\sigma$~\cite{sell_c_sigma}, and (ii) in the use of auto-tuners such as Morpheus-Oracle\cite{morpheus-oracle} and SMAT\cite{smat} to automatically determine the optimal format to use for a computation.

A few recent works have examined the performance of different sparse formats on AArch64 targets.
In particular on A64FX processors, the SELL-C-$\sigma$ matrix storage has been shown to achieve performance and memory-bandwidth saturation superior to the standard \gls{csr} format, reaching performance on par with NVIDIA's V100 GPUs for large, memory-bound SpMV datasets~\cite{alappat2022execution}. An optimised variation of the CSR format---``Bitmap-based CSR (BCSR)''---that extracts edge information more efficiently and has a smaller memory footprint than the classical CSR format has also been proposed for the A64FX-based Fugaku supercomputer, enabling it to reach rank~1 in the Graph500 benchmark in November 2020~\cite{nakao2021performance}.
Other variations of the \gls{csr} and \gls{ell} formats, namely aligned CSR (ACSR) and aligned ELL (AELL), respectively, have also been proposed and implemented with Neon instructions for AArch64 targets~\cite{zhang2021performance}.

For sparse matrix multiplications, FPGA vendors such as Intel implemented, among others, the \gls{coo}, \gls{csr} and \gls{dia} formats in their Sparse BLAS libraries as part of the Intel oneAPI Math Kernel Library \cite{fpga_intel_formats}, whereas Xilinx have developed an implementation of the CSC format in their Vitis Sparse libraries\cite{fpga_xilinx_formats} and the \gls{coo} format in their General Matrix Operation (GEMX)\cite{fpga_xilinx_gemx} engine library which provides building blocks for constructing matrix operation accelerators on FPGA. Other work focused on implementing a modified CSR (MCSR) format for \gls{spmv} multiplication in HLS\cite{fpga_mscr}, or customised sparse matrix formats to leverage the available \gls{hbm} on recent generation FPGAs\cite{fpga_sparse_hbm}.

%% file: src/conclusions.tex
\section{Conclusions and Further Work}
In this paper, we have shed light on the challenges implementing prototypes of the SpMV kernels for the three sparse storage formats COO, CSR and DIA on the two emerging architectures AArch64 CPUs and FPGAs. Optimising the three kernels on the respective architectures, we highlight the performance advantages of individual approaches and show that our Morpheus implementations are competitive. Moreover, we describe potential integration targets of our prototypes to be implemented in the Morpheus library. While our results prove performant implementations especially on AArch64 CPUs with SVE and compared to ARMPL, accelerators such as FPGAs exhibit larger performance portability gaps compared to architectures that do not build on a host-device model.

In terms of future work, a full integration of the FPGA prototype as novel backend to Morpheus is of interest but will require further engineering around abstractions for memory management, build process integration, smart container/layer translations due to restricted availability of dynamic memory and data type support. As vendors such as Intel and AMD-Xilinx have come up with architecture specific implementations of HPCG, an evaluation of these against the Morpheus-HPCG version will be possible, especially with AMD-Xilinx's closely to the HPCG problem size generation tied CSR implentation on FPGA. Moreover, the FPGA prototypes of the storage formats could benefit from implementation on the newest generation of AMD-Xilinx FPGAs, the Versal ACAP.

%- fpga backend integration into morpheus
%- once xilinx backend ported to morpheus: then possible to compare moprheus hpcg vs xilinx hpcg

While our focus in this paper is on the three core sparse matrix storage formats supported by Morpheus i.e. COO, CSR and DIA, there exist a plenitude of other storage formats such as ELL, HYB or HDC that are widely used and from which the Morpheus extensions on AArch64 CPUs and FPGAs could provide further benefit to HPC users.

%- More formats

In terms of theoretical evaluation, future work includes the application of a roofline model to understand and validate the performance of our implementations and compare theoretical performance on our target systems to achieved performance.

Of interest is also the multi-threaded (OpenMP) implementation of the current formats with \gls{sve} intrinsics and benchmark on other Arm systems. Finally, extending \emph{Morpheus} to support a dynamic selection and dispatch mechanism that adapts to the optimal algorithm given a sparsity pattern could be beneficial.
%  Finally, a theoretical evaluation, based on Berkeley’s Roofline model demonstrates that our implementation is near optimally tuned on the Xilinx Alveo U280.
% We applied the Roofline Model to the HPCG implementation on the Xilinx Alveo U280 and underlying hardware, and demonstrated that performance is near-optimal.
% The Roofline model [21,22] is a visually-intuitive method to understand the performance of a given kernel based on a bound and bottleneck analysis approach. The Roofline model characterizes a kernel’s performance in billions of instructions (GFLOPS, y-axis) as a function of its operational intensity (OI, x-axis). We use Operational Intensity as the x-axis and, given that our kernel performs only double operations, use billions of float operations per second (GFLOPS) as the y-axis. Operational Intensity is defined as instructions per byte of memory traffic, which measures traffic between the caches and memory.

%- roofline model for theoretical maximum vs achieved performance
% \textcolor{red}{- Other ARM systems}

% \textcolor{red}{- Multi-threaded (Arm)}

% \textcolor{red}{- algorithm/implementation selection}

%% file: src/acknowledgments.tex
\section*{Acknowledgment}
This research is part of the EPSRC project ASiMoV (EP/S005072/1). We used the Isambard~2 UK National Tier-2 HPC Service (http://gw4.ac.uk/isambard) operated by GW4 and the UK Met Office, and funded by EPSRC (EP/T022078/1). We also acknowledge the ExCALIBUR H\&ES FPGA testbed and AMD Xilinx HACC program for access to compute resource used in this work.